\documentclass[a4paper,11pt]{article}
\pdfoutput=1 
             
\usepackage{HGTDaltiroc-defs}

\usepackage{./latex/jinstpub} 

\usepackage{multirow}
\usepackage{float}
\usepackage{rotating}
\usepackage{mathrsfs}
\usepackage{booktabs}
\usepackage{placeins}
\usepackage{amssymb}
\usepackage{amsmath}
\usepackage{upgreek}
\usepackage{siunitx}
\usepackage{subfig}

\usepackage{graphicx}
\usepackage{HGTDaltiroc-defs}
\newbox{\bigpicturebox}
\usepackage{lineno}

\title{\boldmath Performance of a Front End prototype ASIC for picosecond precision time measurements with LGAD sensors}

\author[a]{C. Agapopoulou,}
\author[b]{S. Blin,}
\author[a]{A. Blot,}
\author[c]{L. Castillo Garcia,}
\author[c]{M. Chmeissani,}
\author[b]{S. Conforti di Lorenzo,}
\author[b]{C. de La Taille,}
\author[b]{P. Dinaucourt,}
\author[a]{A. Fallou,}
\author[c]{J. Garcia Rodriguez,} 
\author[c]{V. Gkougkousis,}
\author[c]{C. Grieco,}
\author[c,d]{S. Grinstein,}
\author[e]{S. Guindon,} 
\author[a]{N. Makovec,}
\author[b]{G. Martin-Chassard,}
\author[f]{G. Pellegrini,}
\author[e]{A. Rummler,}
\author[a]{S. Sacerdoti,}
\author[b]{N. Seguin Moreau,}
\author[a]{L. Serin,}
\author[g]{A. Tricoli}

\affiliation[a]{IJCLab, Univ. Paris-Sud, CNRS/IN2P3, Universit\'e Paris-Saclay, Orsay, France}
\affiliation[b]{Omega, CNRS, Ecole Polytechnique, Palaiseau, France }
\affiliation[c]{Institut de F\'isica d'Altes Energies (IFAE), Barcelona, Spain }
\affiliation[d]{Instituci\'o Catalana de Recerca i Estudis Avan\c cats (ICREA), Barcelona, Spain}
\affiliation[e]{CERN, Geneva, Switzerland}
\affiliation[f]{Centro Nacional de Microelectronica (CNM-IMB-CSIC), Campus UAB, Barcelona, Spain}
\affiliation[g]{Physics Department, Brookhaven National Laboratory, Upton NY, United States of America}

\emailAdd{christina.agapopoulou@cern.ch}

\abstract{

For the High-Luminosity phase of LHC, the ATLAS experiment is proposing the addition of a High Granularity Timing Detector (HGTD) in the forward region, to mitigate the effects of the increased pile-up. The chosen detection technology is Low Gain Avalanche Detector (LGAD) silicon sensors that can provide an excellent timing resolution below 50 ps. The front-end read-out ASIC must exploit the large signal derivative and small noise provided by the sensor, while keeping low power consumption. 
This paper presents the results on the first prototype of a front-end ASIC, named ALTIROC0, which contains the analog stages (preamplifier and discriminator) of the read-out chip. The ASIC was characterised both alone and as part of a module with a 2$\times$2 LGAD array of 1.1$\times$1.1 mm$^2$ pads bump-bonded to it. 
The various contributions of the electronics to the time resolution were investigated in test-bench measurements with a calibration setup. Both when the ASIC is alone or with a bump-bonded sensor, the jitter of the ASIC is better than 20 ps for an injected charge of 10 fC. The time walk effect, which arises from the different preamplifier response for various injected charges, can be corrected up to 10 ps using a Time Over Threshold measurement. The combined performance of the ASIC and the LGAD sensor, which was measured during a beam test campaign in October 2018 with pions of 120 GeV energy at the CERN SPS, is around 40 ps for all measured modules. All tested modules show good efficiency and time resolution uniformity. 
}

\begin{document}
\maketitle
\newpage

\section{Introduction}
\label{sec:intro}
The High Luminosity (HL) phase of the Large Hadron Collider (LHC), to begin in 2027, is expected to deliver instantaneous luminosities more than three times higher than the ones reached during the Run II period. This implies an increase in the average number of collisions per bunch crossing, to around 200. 
In such conditions, pile-up mitigation will be an extremely important subject for the ATLAS experiment \cite{ATLAS}.
The foreseen new generation of pixel detectors, with a reduced pixel size compared to the existing tracker, will manage to keep an excellent track reconstruction performance \cite{ITk}. However, for tracks in the forward region of the detector, the resolution of the vertex longitudinal position will not be as good as in the central region, and tracks coming from different collisions will not always be correctly paired to their corresponding vertices.
The effect of pile-up can be mitigated if an accurate time measurement is combined with the track longitudinal impact parameter, since these characteristics are orthogonal to each other. In this way, pile-up tracks that come from vertices that are very close in distance to the primary vertex, but separated in time, can be removed.
   
In order to implement this concept, the ATLAS experiment is proposing a forward timing detector made of Low Gain Avalanche Detectors (LGADs)~\cite{LGAD}, called the High Granularity Timing Detector (HGTD). The goal is to provide a 50~ps time resolution per track at the level of a minimum ionising particle (MIP), with a layout that guarantees on average 2 or 3 hits per track~\cite{HGTD-TP}; this time resolution should be maintained up to a neutron equivalent fluence of about 2.5$\times$10$^{15}$ n$_{\mathrm{eq}}$/cm$^2$. 
The LGAD sensors have been shown to have low Landau noise, and to be capable of providing a moderate gain \cite{LGAD}. Their development is ongoing to achieve the optimal performance and desired radiation hardness in the framework of the HGTD.

The time resolution is strongly linked to the front-end analog performance, which makes the read-out ASIC a very challenging circuit to design. The time jitter should be low enough to not deteriorate the sensor performance. The requirements have been set to a jitter smaller than 20 ps for the baseline input charge of 10 fC, together with a negligible impact from time-walk (after correction using the signal amplitude or a Time Over Threshold measurement). 
Moreover, the envisioned circuit should be able to provide a time measurement for charges as low as 4 fC, to cope with the reduction of the sensor gain due to irradiation.

A first ASIC, called ALTIROC0 (ATLAS LGAD Time Read Out Chip) has been designed containing the amplifier and the discriminator stages of the final chip. A first version of this prototype has already been studied \cite{ALT0}, and, in this paper, results from an improved second iteration are discussed.  Firstly, a chapter describing some considerations about time resolution is presented, after which the ASIC design is described. Details of the prototype devices used for the purposes of this paper can be found in section \ref{sec:readout_board}. Section \ref{sec:performace_asic} presents test bench measurements of the ASIC. Finally, combined sensor+ASIC results in laboratory and test-beam are discussed in sections~\ref{sec:performance_lgad} and \ref{sec:performance_testbeam}, respectively.

\section{Time resolution consideration}
\label{sec:preamp_design}
The jitter due to electronics noise is often modelled as:
\begin{equation}
  \sigma_{\mathrm{jitter}}= \frac{N}{dV/dt} \sim \frac{t_{rise}}{A/N}
  \label{eq:jitter}
\end{equation}
where N is the noise and $dV/dt$ the slope of the signal pulse of amplitude A and rise-time $t_{rise}$.
Since the noise scales with the bandwidth (BW) as $\sqrt{BW}$, while the signal slope grows with the amplitude as $A\times BW$, the most common timing optimisations rely on using the fastest preamplifier.

Many timing measurements in testbeam have been carried out with broadband amplifiers, which are voltage sensitive amplifiers with a 50~$\Omega$ input impedance. Some prefer to use a trans-impedance configuration and timing optimisation has been published for such configuration~\cite{TestBeamPaper, nicolo}.
However, the preamplifier speed becomes less crucial when dealing with Si or LGAD sensors, because of their current duration (not negligible compared to the preamplifier rise-time) and the sensor capacitive impedance.

\begin{figure}[htbp]
\centering
\subfloat[]{\includegraphics[width=0.6\textwidth]{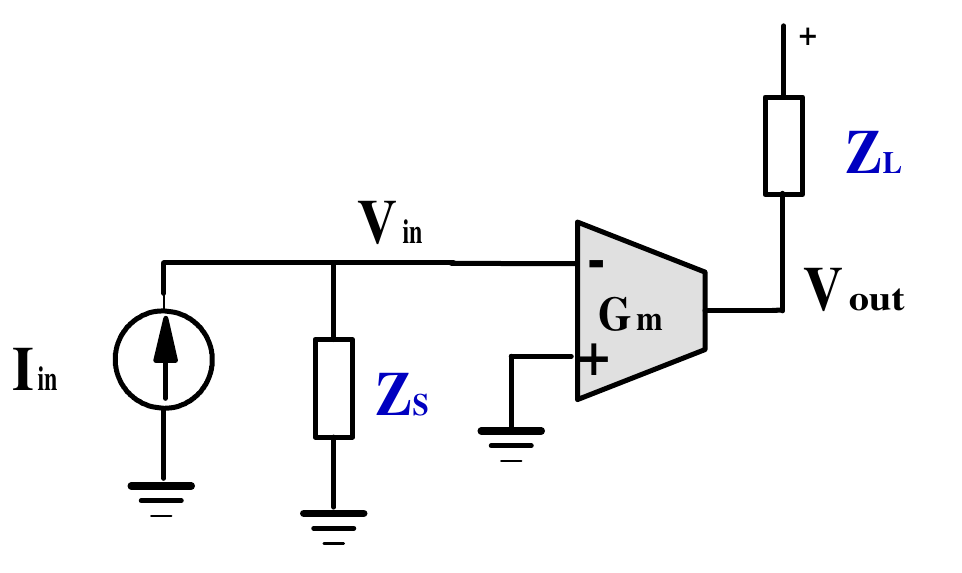}}
\caption{Simple Voltage sensitive amplifier configuration.}
\label{fig:preamp_simple}
\end{figure}

The schematics of a simplified voltage sensitive amplifier configuration are presented in Figure~\ref{fig:preamp_simple}. In such configuration, the jitter can be easily calculated assuming that the detector is a constant current source $I_{in}$ with a duration time of $t_{dur}$. The corresponding input charge $ Q_{in}$ is then equal to $I_{in} \times t_{dur}$.
$I_{in}$ is converted into an input voltage ($V_{in}$) through the overall input impedance. The latter is given by the sensor input impedance $Z_s=1/j\omega C_d$ (C$_d$ is the total detector capacitance) in parallel with the preamplifier input impedance $R_{in}$.
At high frequency and for C$_d \sim$ 1-10 pF, the input impedance is dominated by the detector capacitance, and therefore, the signal is integrated in C$_d$. At this high-speed domain, the maximum voltage is reached before the preamplifier feedback network reacts and drains out the charge from the input capacitance. The input voltage is then given by $V_{in} = \int I_{in} (t)/C_d dt = Q_{in}/C_d$. The preamplifier output voltage is $V_{out}=g_m Z_L V_{in}$, where $g_m$ represents the transconductance of the transistor and Z$_L$ the preamplifier load impedance. The output signal would reach its maximum in the input pulse drift duration time (t$_{dur}$) if the preamplifier was infinitely fast.  With a real preamplifier, where the output signal is the convolution of the input current and the preamplifier response, a convenient approximation to take into account its speed is given by the quadratic sum of the t$_{dur}$ and the preamplifier rise-time (t$_{r_{pa}}$):  $\sqrt{t_{dur}^2+t_{r_{pa}}^2}$. If, instead of a constant current, the LGAD's trapezoidal signal is considered, the result is quite similar, but the Full Width at Half Maximum (FWHM) of the detector current pulse, t$_{FWHM}$, is used instead of t$_{dur}$.

The voltage RMS (V$_n$) at the preamplifier output and the signal slope (dV/dt) are then given by: 
\begin{equation}
 V_n= G_{pa} \times e_n \sqrt{\pi*BW/2} \sim \frac{G_{pa} \times e_n}{\sqrt{2 t_{r_{pa}}}}
 \hspace{0.2cm} {\rm  and } \hspace{0.2cm}  \frac{dV}{dt} = \frac{G_{pa} Q_{in}}{C_d \sqrt{t_{r_{pa}}^2+t_{d}^2}}
\end{equation}
where G$_{pa}$ is the preamplifier gain, e$_n$ the noise spectral density, and t$_{d}$ is either the t$_{dur}$ in the case of a constant current source or the t$_{FWHM}$ in the case of an LGAD pulse.
Combining all the terms results in the following formula for the jitter:
\begin{equation}
  \sigma_{jitter}= \frac{e_n C_d}{Q_{in}} \sqrt{\frac{t_{r_{pa}}^2+t_{d}^2}{2 t_{r_{pa}}}}
  \label{ec:jitter}
\end{equation}
It can be seen that the condition to minimize the jitter is to match the preamplifier rise-time to the t$_d$ : $t_{r_{pa}} = t_{d}$, thus reducing the jitter formula to:

\begin{equation}
\sigma_{jitter}= \frac{e_n C_d}{Q_{in}} \sqrt{t_{d}}. 
\end{equation}

However, this dependence is not very strong. For instance, for a sensor drift-time of 600~ps, if the preamplifier rise-time is reduced or increased by a factor of 2 compared to the optimal matching value, the jitter would deteriorate by just about 12\%.   
Given these considerations, in order to minimize the jitter, the sensor should have a small capacitance, a short pulse duration, and be capable of providing a large charge.
The ATLAS baseline choice is LGADs with a pixel size of 1.3 $\times$ 1.3 mm$^2$ and a \SI{50}{\micro\metre} active thickness, to be operated with a starting (minimum) collected charge of at least 10 (4) fC, i.e a gain of 20 (8).\footnote{For \SI{50}{\micro\metre} thick LGADs the gain is roughly twice the injected charge.} The electronics noise e$_n$ is largely determined by the current (I$_d$) that can be flown in the preamplifier input transistor, as explained in the next section.

\section{ALTIROC0 design}
\label{sec:altiroc0_design}

\subsection{Preamplifier} 

\begin{figure}[htbp]
\centering
\subfloat[]{\includegraphics[width=0.6\textwidth]{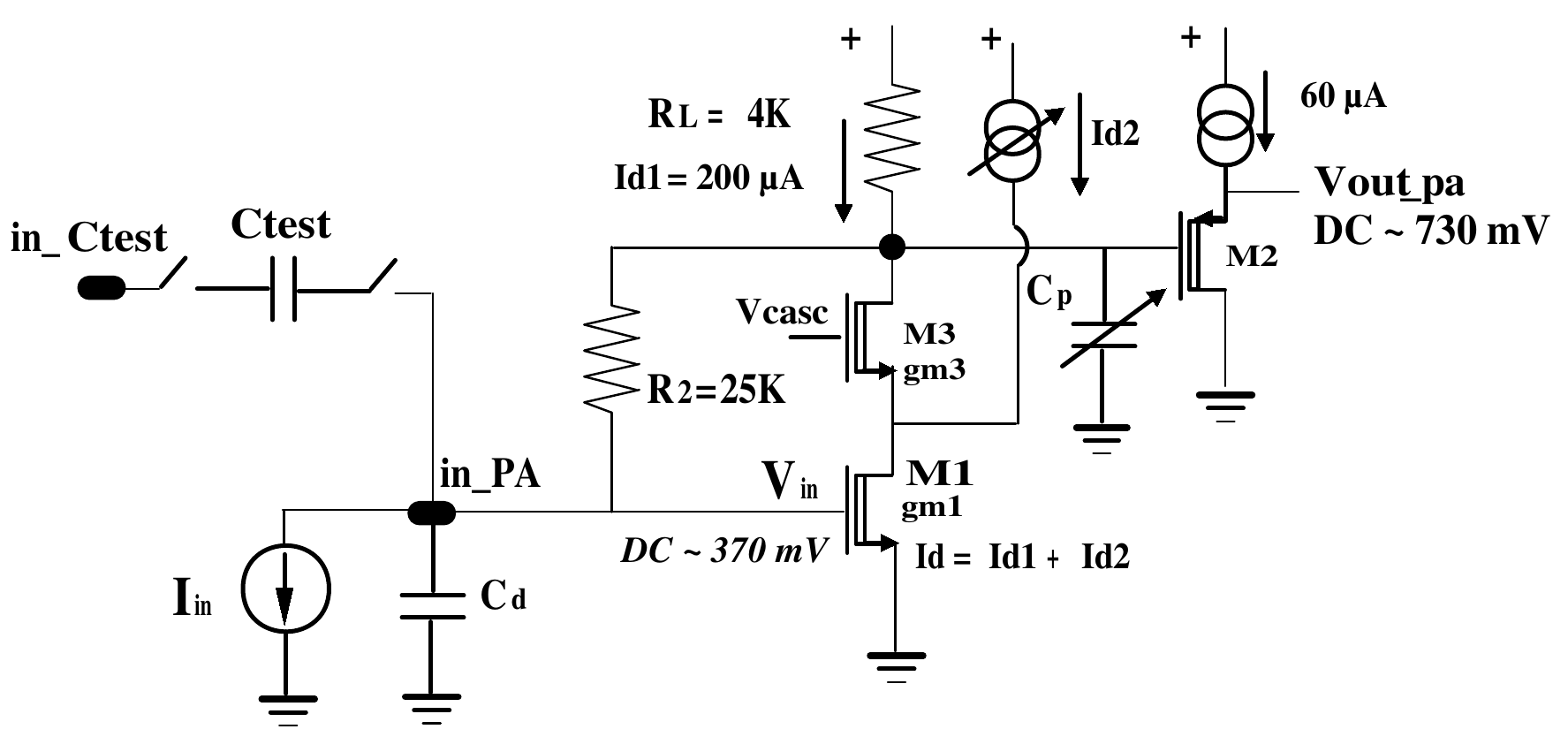}\label{fig:preamp_volt}}

\caption{Architecture of the preamplifier}
\label{fig:preamp_schem}
\end{figure}

The ALTIROC0 preamplifier, shown in Figure~\ref{fig:preamp_schem}, is a voltage preamplifier built around a cascoded common source configuration (M1) followed by a voltage follower (M2). The R$_2$ resistor ensures the biasing of the preamplifier input and can be used to adjust the fall time of the preamplifier output. 

Given that the preamplifier is voltage sensitive, the detector capacitance is a key ingredient to calculate the input voltage for a given input charge. An input charge Q$_{in}$ gives an input voltage $V_{in}$ equal to $Q_{in}/C_d$. The voltage output of the preamplifier is given by the following expression: 
\begin{equation}
Vout= G_{pa}\times V_{in} = G_{pa}\times Q_{in}/C_d \label{eq:vout}
\end{equation}
The gain of the preamplifier $G_{pa}$ is, to first order, given by g$_{m1} \times R_L$, where $g_{m1}$ is the transconductance of the input transistor. In weak inversion, the transconductance is given by: 
\begin{equation}
    g_{m1} = q\times I_{d}/2kT
    \label{eq:transconductance}
\end{equation} where q=1.6 10$^{-19}$ C, i.e. approximately $20 \times I_{d}$ at room temperature. 
The spectral density of the input transistor is equal to:
\begin{equation}
e_n = \sqrt{2kT/g_{m1}}.
\label{eq:noise_density}
\end{equation}  
As both gain and noise depend on the current that flows in the input transistor, the drain current I$_{d}$ is made of two current sources: I$_{d1}$ is a fixed current source of \SI{200}{\micro\ampere} while I$_{d2}$ can be tuned with an external resistor.\footnote{In the second iteration of this ASIC, the current source I$_{d2}$ can be tuned by slow control parameters from 0 to \SI{850}{\micro\ampere} with a DAC.} Simulations have shown that increasing I$_{d2}$ beyond \SI{600}{\micro\ampere} adds little gain, as the transistor is no more in weak inversion mode.

To compensate for the rise-time of the LGAD sensor becoming smaller when irradiated, the preamplifier rise-time is tuneable. This is done through the pole capacitance C$_p$ that can be adjusted through slow control (from 0 to 175 fF), allowing to set a preamplifier rise-time between 300 ps and 1 ns (bandwidth between 350 MHz to 1 GHz).

The input impedance R$_{in}$ is given by the R$_2$ resistance divided by the open-loop gain of the preamplifier. The value of R$_{in}$ depends therefore also on the drain current I$_d$. For I$_d$= \SI{300}{\micro\ampere} and R$_2$=25k$\Omega$, the input impedance is around 1.6k$\Omega$. The preamplifier fall time depends on the time constant, which is given by R$_{in}$ multiplied by the total capacitance seen on the preamplifier input (sum of the sensor capacitance and any parasitic capacitance). With 3-4 pF capacitance, this fall time is within the bunch-crossing interval of the HL-LHC, that is 25 ns. However, the width of the discriminator pulse, which is used to correct for the time walk effect (\ref{subsec:discri}), can be slightly longer than this interval, provided a large injected charge and a low threshold. This could disturb the measurements and, therefore, the value of the R$_2$ resistance will be reduced in the next iteration of the ASIC. The resistance R$_2$ is also used to absorb the sensor leakage current I$_{leak}$. This leakage current would induce a drift of the preamplifier output DC voltage by an amount of the order R$_2$ $\times$ I$_{leak}$. The discriminator threshold needs to be corrected accordingly to this shift.
After irradiation at the largest fluence expected at the end of the HGTD lifetime (2.5$\mathrm{\times 10^{15} n_{eq}/cm^{2}}$), the maximal leakage current of the LGAD sensor is estimated to be below \SI{5}{\micro\ampere}.

Finally, in order to inject an accurate calibration charge, a calibration capacitor (C$_{test}$=100 fF), which can be selected by slow control, is also integrated. With a fast voltage step of 100 mV, provided by a pulse generator of picosecond level precision, a very short square pulse with a charge of 10 fC is delivered at the preamplifier input. Such an input signal allows the characterisation of the front-end read-out but does not reproduce the jitter performance with an LGAD pulse, as the signal shape and time duration can not be neglected. For the same input charge, the simulation predicts a jitter larger by a factor 1.65 when using as input the LGAD signal instead of the calibration signal. 

\subsection{Discriminator}\label{subsec:discri}
The measurement of the Time of Arrival (TOA) of the particles is performed by a discriminator that follows the preamplifier.  The discriminator uses a programmable threshold which induces a dependence of the time measurement on the signal's peak height, an effect called time walk. The measurement of the time of the rising edge of the discriminator pulse provides the TOA while that of the falling edge, combined with the TOA, provides the Time Over Threshold (TOT). The TOT is an estimate of the pulse amplitude and can be used to correct for the time 
walk effect.\footnote{A constant fraction discriminator was also included as an alternative in some channels of a first version of the chip but the performance was worse than applying a TOT-based time walk correction. Finally, this alternative was removed from the design of the current prototype.} To ensure a jitter smaller than 10 ps at large signals, the discriminator is built around a high speed leading edge architecture. Two differential stages with small input transistors are used to ensure a large gain and a large bandwidth (around 0.7 GHz).  The discriminator threshold (V$_{th}$) is set by an external 10-bit DAC, common to all channels.

\subsection{Layout} 

A prototype with 8 channels has been designed in CMOS 130 nm by OMEGA microelectronics.\footnote{https://portail.polytechnique.edu/omega/en/presentation/omega-brief} It integrates four voltage sensitive preamplifier channels and 4 pseudo-trans-impedance preamplifier channels which are not discussed in this paper. Each channel is made of a preamplifier followed by a discriminator.
The design of the chip includes bump bonding pads on each input and also on ground pads. The size of the chip is 3.4 mm x 3.4 mm to accommodate the bump bonding to a 2$\times$2 sensor array with a 1.1 mm $\times$ 1.1 mm pad size. 

\section{ALTIROC0 devices}
\label{sec:readout_board}
This section presents the devices that were used to characterise the performance of the ALTIROC0 ASIC. Dedicated read-out boards were produced on which the ASIC was wire-bonded, either alone, or bump-bonded to an LGAD sensor. In the latter case, the ASIC + sensor system is referred to as a bare module.

\subsection{Read-out boards}
\begin{figure}[H]
\centering

\begin{minipage}[b][6cm][s]{0.6\textwidth}
\centering
\sbox{\bigpicturebox}{
  \subfloat[]{\label{fig:standardBoard}\includegraphics[width=\textwidth,height=6cm]{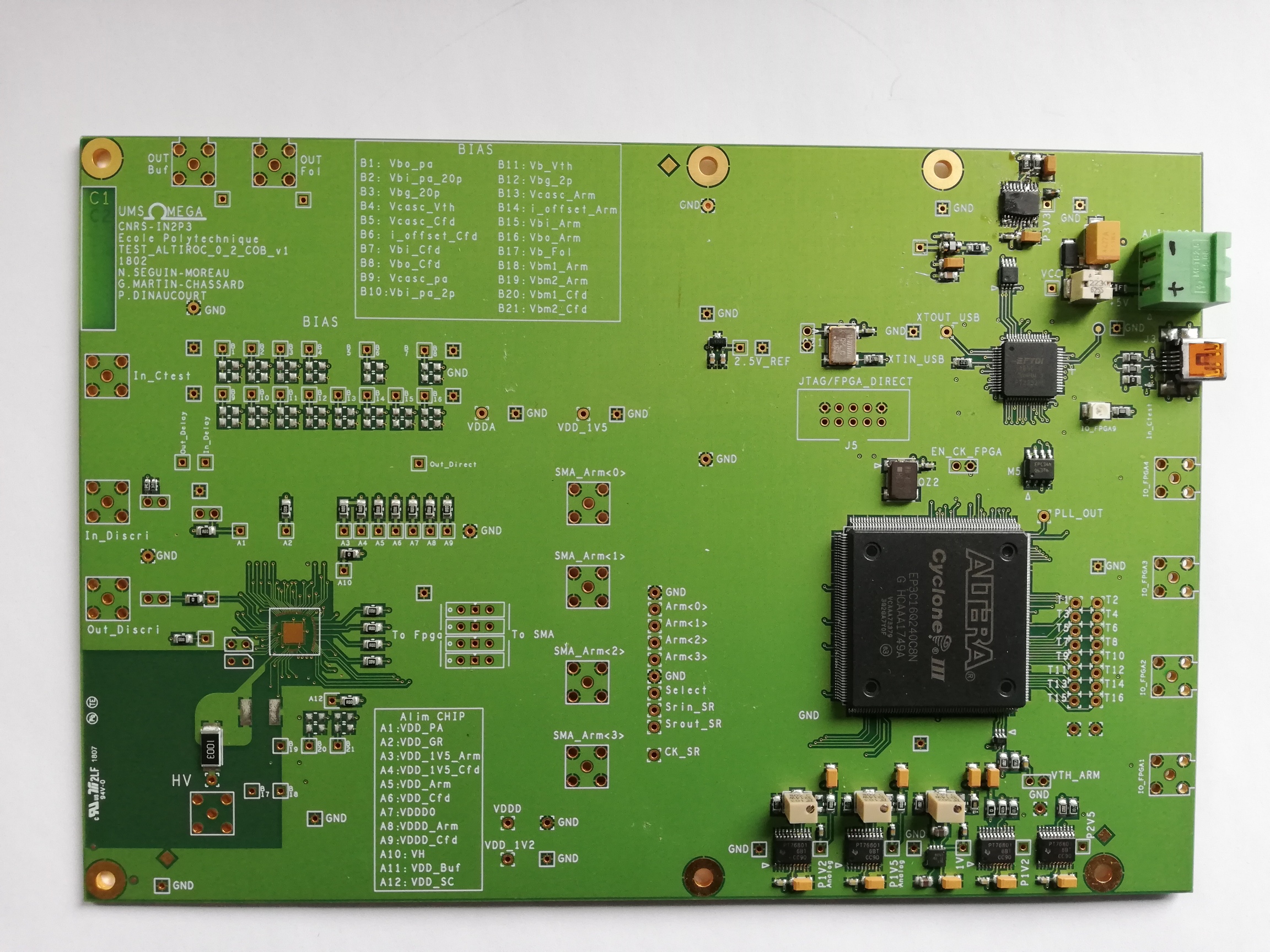}}}

\usebox{\bigpicturebox}
\end{minipage}\hfill
\begin{minipage}[b][6cm][s]{0.3\textwidth}
\centering
\subfloat[]{\label{fig:boardStand_zoom}\includegraphics[width=0.6\textwidth,height=2.5cm]{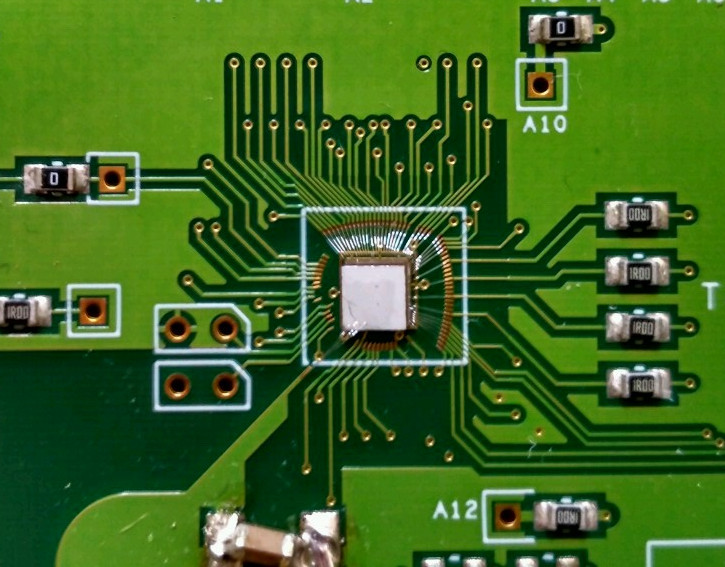}}

\subfloat[]{\label{fig:LshapeBoard}\includegraphics[width=0.6\textwidth,height=2.5cm]{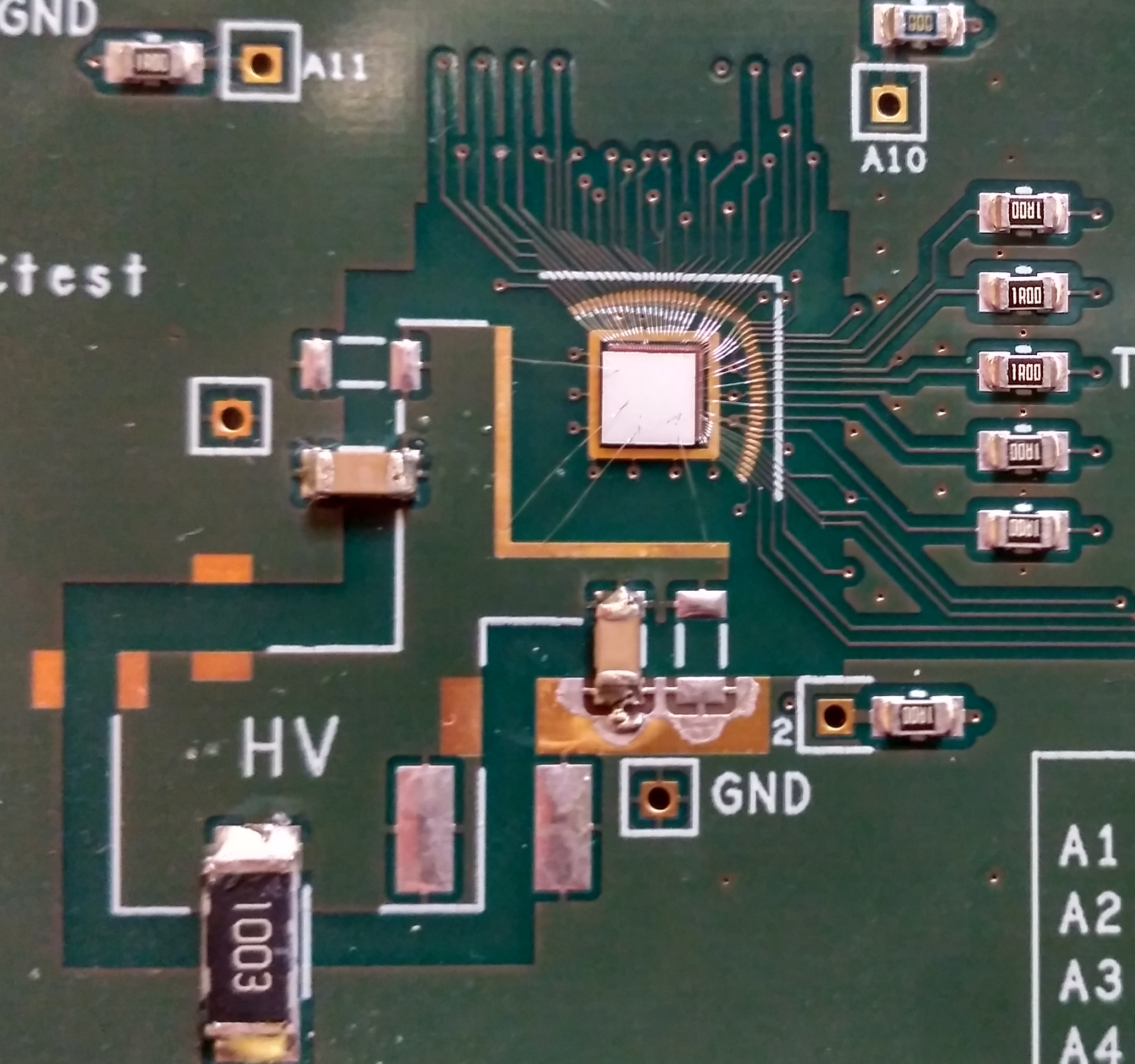}}
\end{minipage}

\caption{(a) Photograph of a standard ALTIROC0 board. (b) Zoom of figure (a) on the flip-chip consisting of an ALTIROC0 ASIC bump bonded to a 2$\times$2 LGAD sensor array. (c) Zoom on the flip-chip area of a modified L-shape HV pad board. }
\end{figure}

A picture of  the custom board used to characterise the ASIC is shown in Figure~\ref{fig:standardBoard}. The board is equipped with a standard Field Programmable Gate Array (FPGA) used to load the slow control parameters. The four discriminator outputs can be read-out directly on SubMiniature version A (SMA) connectors. A dedicated probe is available on an SMA connector to read the preamplifier output after a second stage amplifier/shaper. The channel(s) to be read-out is selected through slow control. Finally, an additional SMA connector is used to inject the calibration pulse.

External capacitors can be soldered on the board to mimic the LGAD C$_d$ at the preamplifier input when a sensor is not bump-bonded to the ASIC. In case the bare module is mounted, the preamplifier input is directly connected to the sensor, and therefore, the capacitance at the preamplifier input is equal to the C$_d$ plus the parasitic capacitance of the ASIC.

Two versions of the custom read-out boards were produced to investigate the time-over-threshold issue observed when an LGAD sensor is connected to the input of the ASIC, that is discussed in section \ref{subsec:totProblem}. The second version, as seen in Figure \ref{fig:LshapeBoard}, has an L-shaped High Voltage (HV) pad that allows for multiple HV wire bonds to be connected far from each other, minimising any possible inductance to the HV decoupling capacitor.

\subsection{ASIC-sensor interconnection}\label{subsec:bumpdond}

The interconnection of the sensor to the front-end chip is a critical procedure of the device assembly process.
Each sensor channel is DC-coupled to the corresponding read-out channel on the ASIC through a small electrically conductive bump ball, that is put in place through a hybridisation process called \textit{bump-bonding}. 
Most of the devices presented here were assembled using SnAg solder bumps, which is the baseline HGTD assembly process. However, in one device, gold bumps were used.

Solder bump-bonding consists of three steps. First, under-bump metallization  (UBM) is deposited on both sensor and ASIC pads. Then solder bumps are deposited on the ASIC, and finally, the sensor and ASIC channels are interconnected. The hybridisation process was done on single tiles, i.e., both sensor and ASIC were already diced before UBM.

The \SI{90}{\micro\metre} wide aluminium pads of the sensor and read-out chip were covered with 4 to \SI{6}{\micro\metre} of NiAu through an auto-catalytic chemical technique. The substrates were inspected and excess of UBM on the edges, if present, was removed. SnAg solder bumps of \SI{80}{\micro\metre} diameter were then deposited on the ASICs with a bump deposition machine. The solder bumps were further reflowed in a dedicated machine in order to improve the placement and the shape uniformity of the bump balls.
Flip-chip was performed with a bonder machine that allows to align, heat and press together the two substrates. After flip-chip, the assemblies were reflowed once again with formic acid. 
In total, eight assemblies were produced following this procedure.

Inspection of the devices was carried out using x-rays to verify the good connectivity of all the bump bonds. The topology of the bumps was found to be mostly cylindrical, with a diameter of about  \SI{90}{\micro\metre} and a height of  \SI{50}{\micro\metre} approximately.

An alternative process using Au bumps has also been developed to assemble one of the modules. With Au bumps, UBM is not needed since the ball bumps can be deposited directly on the aluminium of the front-end pads. An alignment and thermo-compression cycle is used to interconnect the channels of the sensor and ASIC. Studies determined that the bump topology resembled a conical frustum with a base of about \SI{140}{\micro\metre} and a height of  \SI{15}{\micro\metre}.

\subsection{Available devices}
Table \ref{tab:dutlist} lists the Devices Under Test (DUTs) that were available for the measurements performed in this paper. The DUT in this case consists of an ALTIROC0 ASIC, wire bonded to a custom readout board, while an LGAD sensor might also be bump-bonded to the ASIC.
Tests of the performance of the ASIC without the presence of a sensor were performed with DUT A3. A board with a modified L-shaped HV pad was equipped with a 2$\times$2 LGAD sensor array, using SnAg bumps with UBM for the bump-bonding, and characterised with the calibration setup (DUT A4). This device was not available for the October 2018 testbeam campaign.

For the October testbeam campaign, the results of which are presented in section \ref{sec:performance_testbeam}, two ALTIROC0 standard boards were available. Both were equipped with a 2$\times$2 unirradiated sensor array that was bump bonded (section \ref{subsec:bumpdond}) on the ASIC. Both sensor arrays were CNM LGAD with a  \SI{50}{\micro\metre} active thickness and 1.1 $\times$1.1 mm pixel size. The two boards and ASICs were identical. The bump and wire bonding of the two boards were performed in different laboratories; one of them, labelled DUT A1, was assembled in IFAE using SnAg bumps with UBM for the bump-bonding. The second one, labelled DUT A2, was assembled in BNL, while Au bumps without UBM were used for the bump-bonding. In A2, channel 1 was discovered before the testbeam to be disconnected, probably due to a faulty contact of the bump.

\begin{center}
\begin{table}[H]
\caption{List of available DUTs, consisting of an ALTIROC0 ASIC wire bonded to a readout board. }

\centering

\begin{tabular}{|c|c|c|c|c|}
\hline

DUT name & active channels & with an LGAD sensor&  HV pad shape  & testbench/testbeam   \\ \hline
A1 & 4 & yes (SnAg + UBM) & standard & both \\ \hline
A2 & 3 & yes (Au) & standard & testbeam \\ \hline
A3 & 4 & no & standard & calibration \\ \hline
A4 & 4 & yes (SnAg + UBM) & L-shape & calibration \\ \hline

\end{tabular}

\label{tab:dutlist}
\end{table}

\end{center}

\section{ASIC test bench performance}
\label{sec:performace_asic}
 As a first step, the performance of the ASIC alone was evaluated with a calibration injection setup in which the ASIC was  wire-bonded on a dedicated read-out board.

\subsection{Calibration test bench setup}
\label{subsec:setup}
A generator with a picosecond level  precision (Picosecond Pulse Labs model 4600)
is used to generate a step pulse of an accurately defined voltage with a 
70 ps rise-time. This signal is injected through the internal 100 fF capacitor, thus producing a very short square pulse with a very precise injected charge at the preamplifier input. A high-frequency splitter is used to duplicate the injected signal to be also used as a time reference for the time resolution measurement. The generator time resolution has been measured to be about 6 ps. The generator provides also the acquisition trigger, done with a Lecroy oscilloscope having a 20 GSamples/s sampling rate and 2.5 GHz bandwidth. The full waveforms are registered for each trigger and analysed off-line.

\begin{subsection}{Measurements}
Most of the measurements were done with an additional external soldered capacitor (C$_{sold}$) of 2 pF to emulate the sensor capacitance. This value was chosen to match the jitter from calibration measurements of boards with a mounted module (ASIC+sensor), that are presented in section \ref{sec:performance_lgad}.

Charge scans were performed from 5 to 50 fC as the typical charge deposited. As explained previously, the current I$_{d2}$ can be modified by an external resistor. For most measurements, a value of I$_{d2}$=\SI{600}{\micro\ampere} was used, resulting in a total current of I$_d$ = \SI{800}{\micro\ampere}.

\begin{subsubsection}{Pulse properties}

Figure \ref{fig:avDiscri} shows the average discriminator response for different injected charges from 5 to 50 fC: the larger the input charge, the larger is the pulse width and the earlier the pulse time.  The average pulse shape of the preamplifier probe is shown in Figure \ref{fig:avProbe} for various values of soldered capacitance. As expected, the pulse amplitude decreases with the capacitance, while the falling time also becomes longer.

\begin{figure}[htbp]
\centering

\subfloat[]{\includegraphics[width=0.5\textwidth]{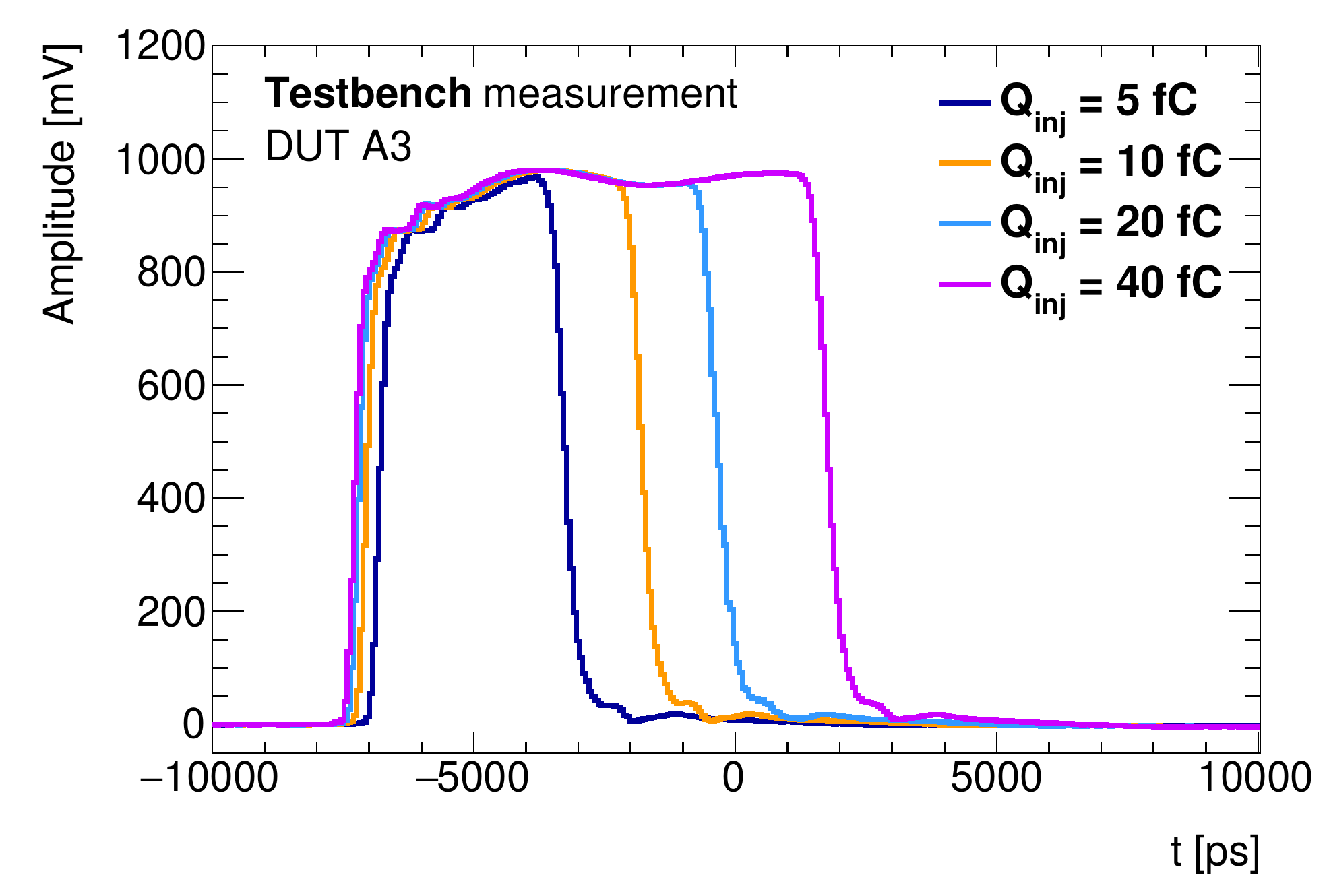}\label{fig:avDiscri}}
\hspace*{0.2cm}
\subfloat[]{\includegraphics[width=0.5\textwidth]{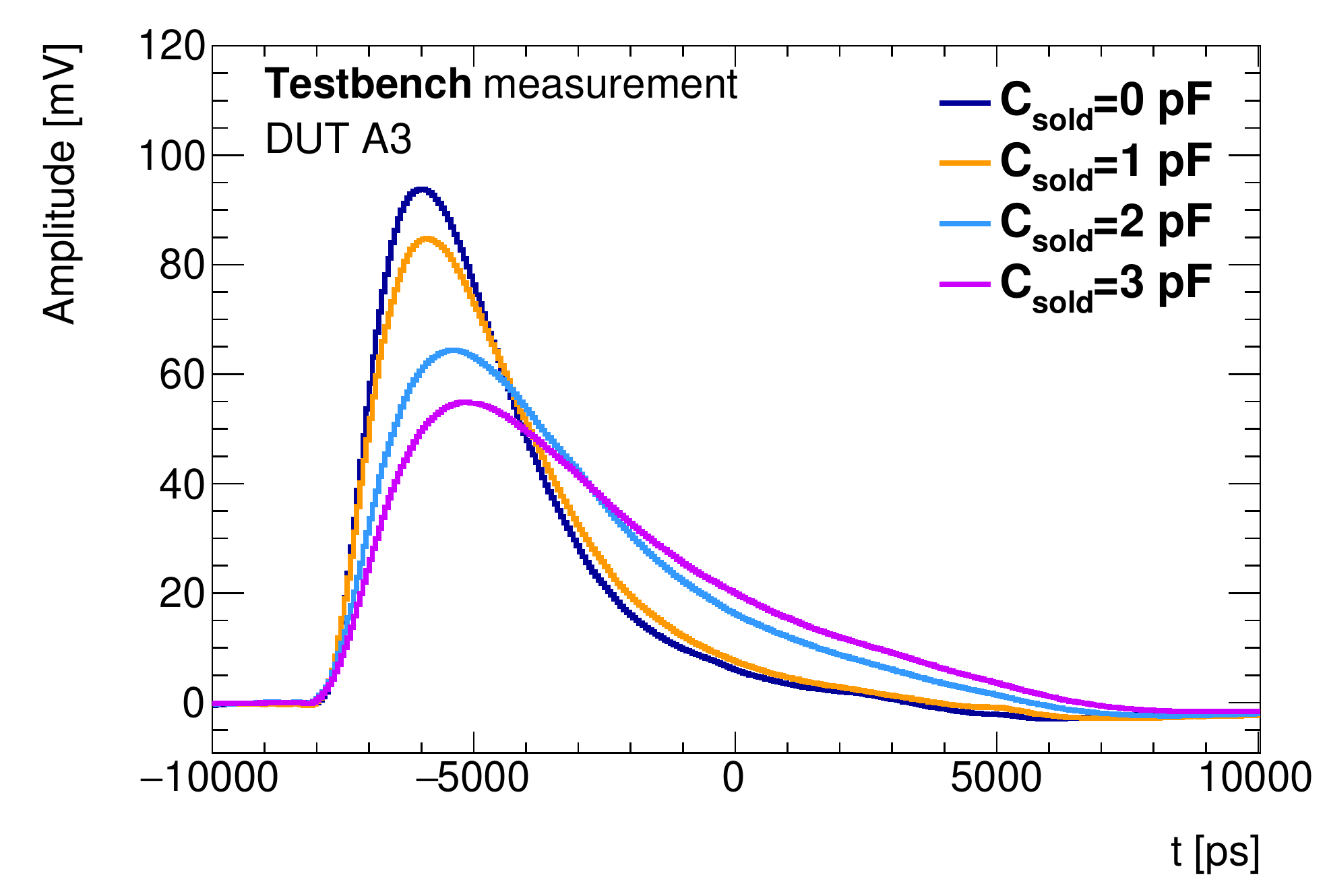}\label{fig:avProbe}}

\caption{(a) Average discriminator pulse shape for Q$_{inj}$=5-40 fC and C$_{sold}$=2 pF. (b) Average probe pulse for C$_{sold}$=0-3 pF and for Q$_{inj}$=10 fC.}
 
\end{figure}

\end{subsubsection}

\begin{subsubsection}{Parasitic capacitance} \label{subsubsec:cpar}
 Apart from the sensor capacitance (or the soldered capacitance in the case of an ASIC alone), there are two additional contributions to the total capacitance to be considered; the parasitic capacitance of the ASIC itself, and the parasitic capacitance of the custom board. Of the two, only the former is relevant to the module performance, since, when the ASIC is bump-bonded to the sensor, the preamplifier input is directly connected to the sensor.
 
 As shown in Eq. \ref{eq:vout}, the total detector capacitance is inversely proportional to the amplitude of the preamplifier output. Under the assumption that $C_d = C_{sensor}+C_{par}$, where $C_{par}$ is the parasitic capacitance, Eq. \ref{eq:vout} can be modified as follows:
\begin{equation}
    \frac{1}{Vout_{pa}} = \frac{C_{sensor}}{G_{pa}*Q_{in}} + \frac{C_{par}}{G_{pa}*Q_{in}}  \label{eq:cpar}
\end{equation}
 
 The contribution of the ASIC to the $C_{par}$ was estimated from a channel whose input had been disconnected from the board, using the amplitude of the preamplifier probe as an estimate of $Vout_{pa}$. It was measured to be 0.8 pF, a value that is expected from simulation.

\end{subsubsection}

\begin{subsubsection}{Jitter}
\label{subsubsec:jitter_asic}

The jitter was calculated from a Gaussian fit to the difference between the discriminator output time and the trigger input signal. For both discriminator and trigger input, the time was measured at the 50\% of the maximum amplitude. Figure \ref{fig:toa1D} demonstrates that the time distribution for a 10 fC input charge is well modelled by a Gaussian with a 13 ps resolution. Figure~\ref{fig:jitterQinj} shows the jitter as a function of the injected charge, for a 2 pF soldered capacitance and a discriminator threshold of 2.5 fC. The trigger time resolution has been quadratically subtracted. The red line corresponds to a fit which follows the theoretical prediction of Eq. \ref{ec:jitter}. The resolution reaches a plateau of 4 ps at high charges.    

The jitter is also shown in Figure \ref{fig:jitterCs} for a 10 fC input charge as a function of the soldered capacitance. As expected, a linear dependence is observed, thus justifying the choice of small area pad sensors with an active thickness of \SI{50}{\micro\metre} for the final detector. 

\begin{figure}[htbp]
\centering

\subfloat[]{\includegraphics[width=0.48\textwidth]{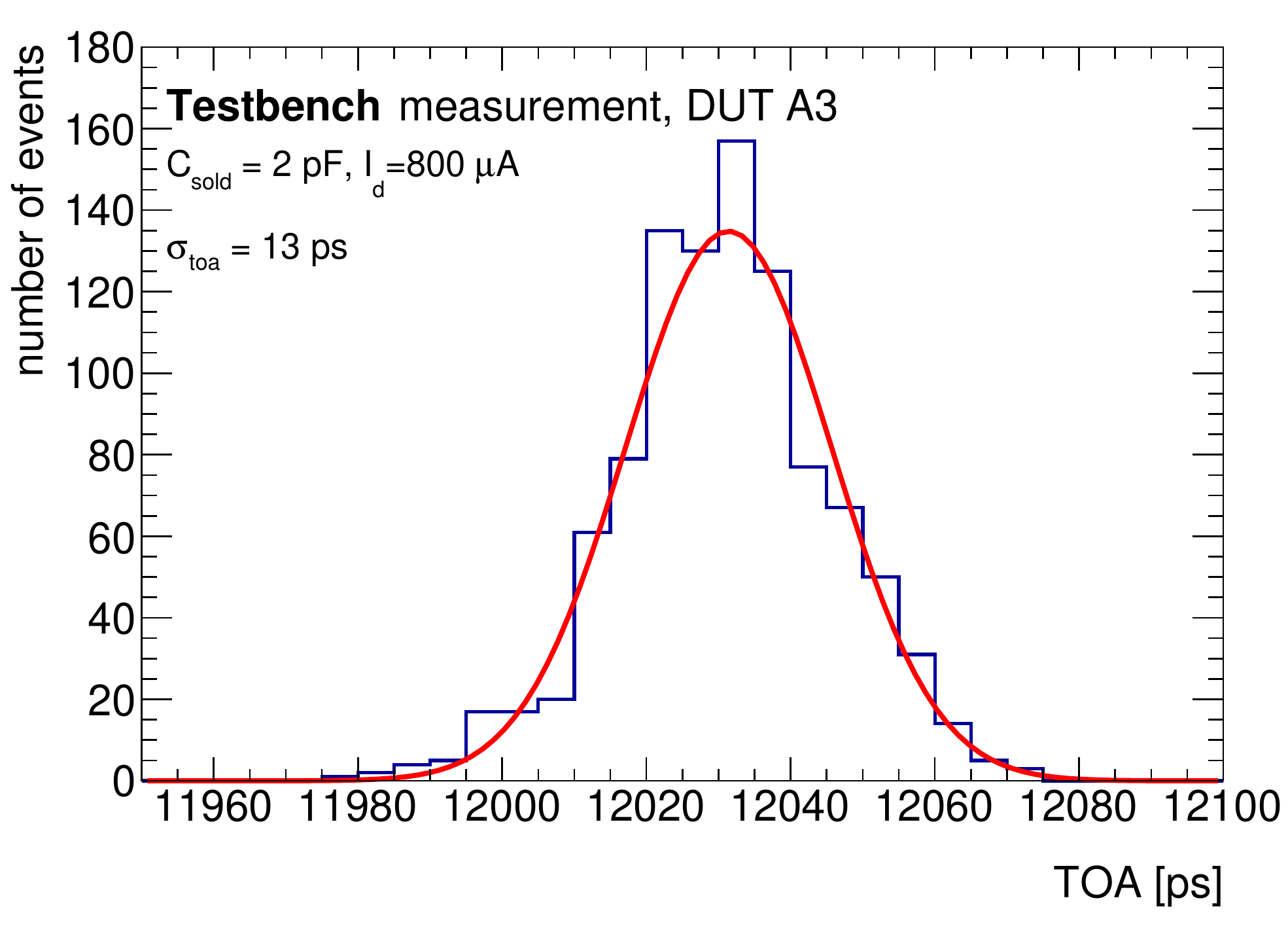}\label{fig:toa1D}}
\hspace*{0.2cm}
\subfloat[]{\includegraphics[width=0.5\textwidth]{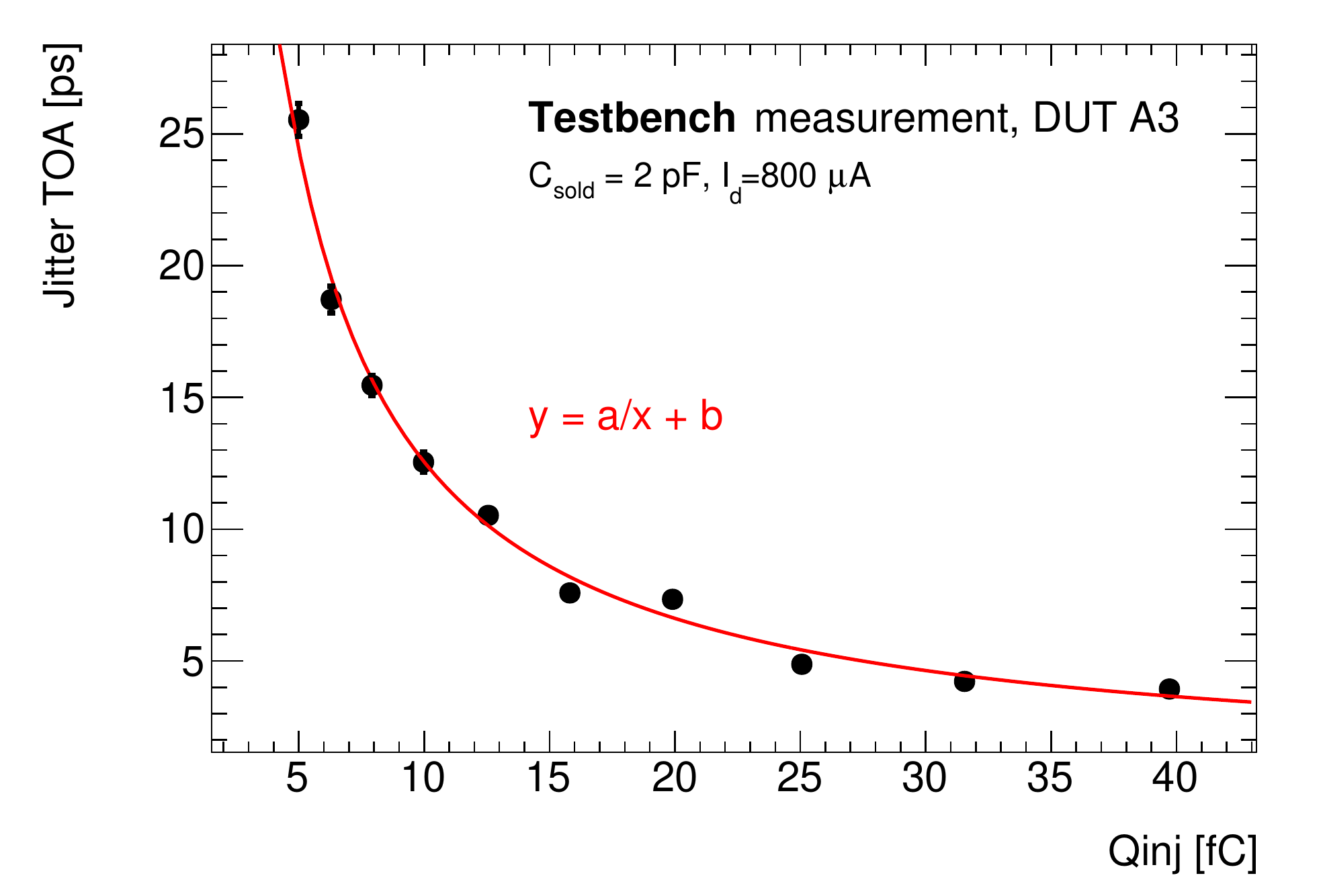}\label{fig:jitterQinj}}

\caption{(a) TOA distribution for Q$_{inj}$ = 10fC. The RMS of the distribution, i.e. the jitter, is found to be 13 ps. The red line corresponds to a Gaussian fit. (b) Jitter as a function of the injected charge. Both measurements have been done with a C$_{sold}$ = 2 pF and a 2.5 fC discriminator threshold.}

\end{figure}

\begin{figure}[htbp]
    \centering
   \includegraphics[width=0.5\textwidth]{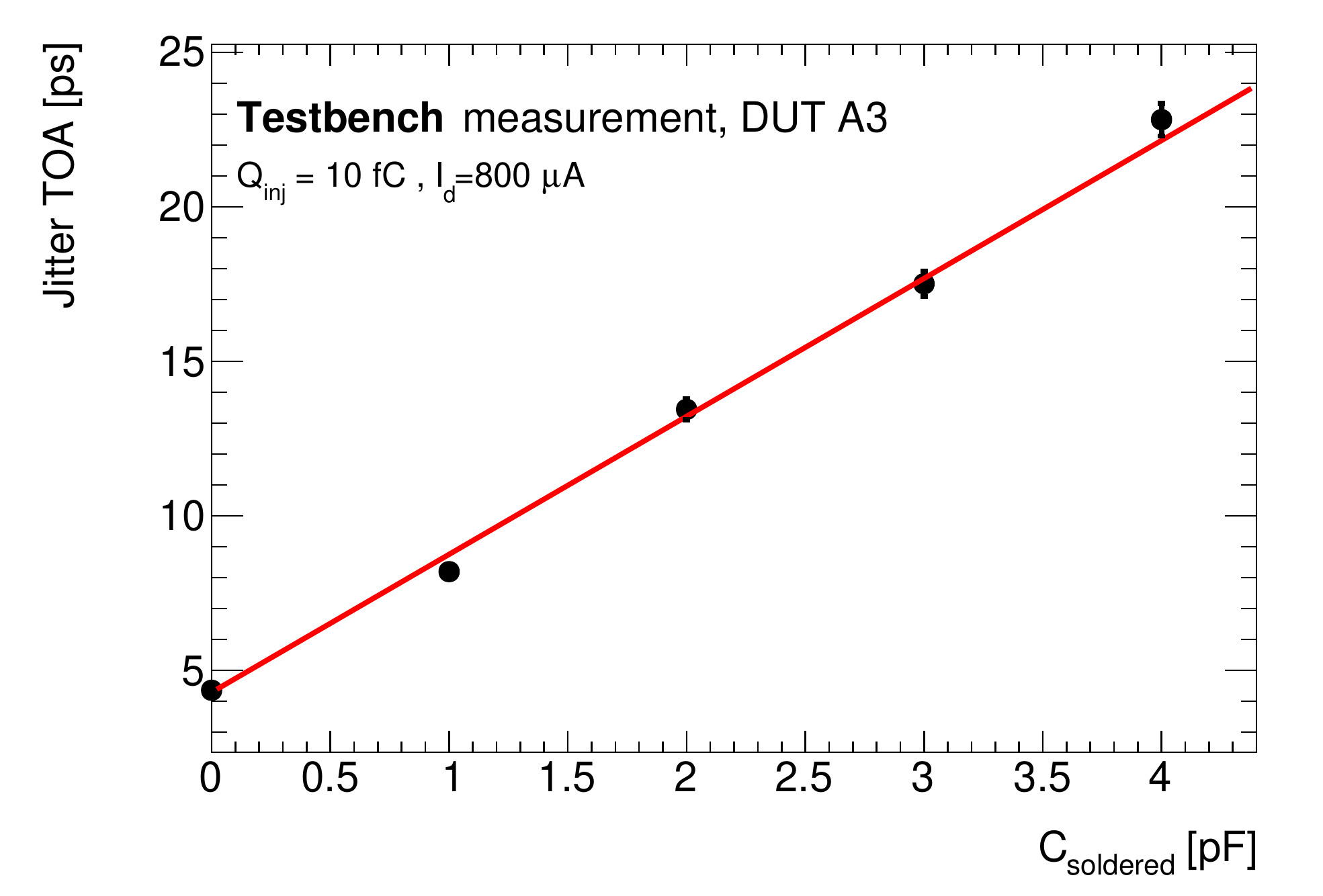}
   
    \caption{Jitter as a function of the soldered capacitance for an injected charge of 10fC. The solid line corresponds to a linear fit.}
   
    \label{fig:jitterCs}
   
\end{figure}

\end{subsubsection}

\begin{subsubsection}{Time walk correction}

The discriminator TOT will be used as an estimate of the input charge to correct for the time walk effect. Figure \ref{fig:tot1D_12fC} shows the TOT distribution for 12 fC input charge. As seen previously for the TOA, the distribution is well modelled by a Gaussian fit with a standard deviation of 120 ps. The correlation of the TOT and the probe amplitude with the input charge is displayed in Figure \ref{fig:totAmp_vs_q}, where it can be seen that the behaviour of these two variables is similar.

The average time of arrival (TOA) as a function of the TOT or the probe amplitude is shown in Figures \ref{fig:toavstot} and \ref{fig:toavsProbe}, respectively, for a soldered capacitance of 2 pF and an injected charge ranging from 5-40 fC. The red line in both figures corresponds to a polynomial fit used to apply the time walk correction.\footnote{A 3$^{rd}$ degree polynomial fit in (1/x) is used.} A time walk of about 500 ps is observed, corresponding to a total bandwidth of 700 MHz for the preamplifier and discriminator. The bottom pad of \ref{fig:toavstot} shows the TOA residuals after correcting for the time walk using the TOT. They are calculated to be in a peak-to-peak range of 40 ps, while a better performance of 20 ps is achieved using the probe amplitude, presented in the bottom pad of Figure \ref{fig:toavsProbe}. In both cases, assuming a pessimistic uniform distribution of the peak-to-peak residual, the achieved residual RMS is $\leq$ 10 ps. This value is consistent with the requirements of the time-walk correction performance for the HGTD.

\begin{figure}[htbp]
\centering
\subfloat[]{\includegraphics[width=0.5\textwidth]{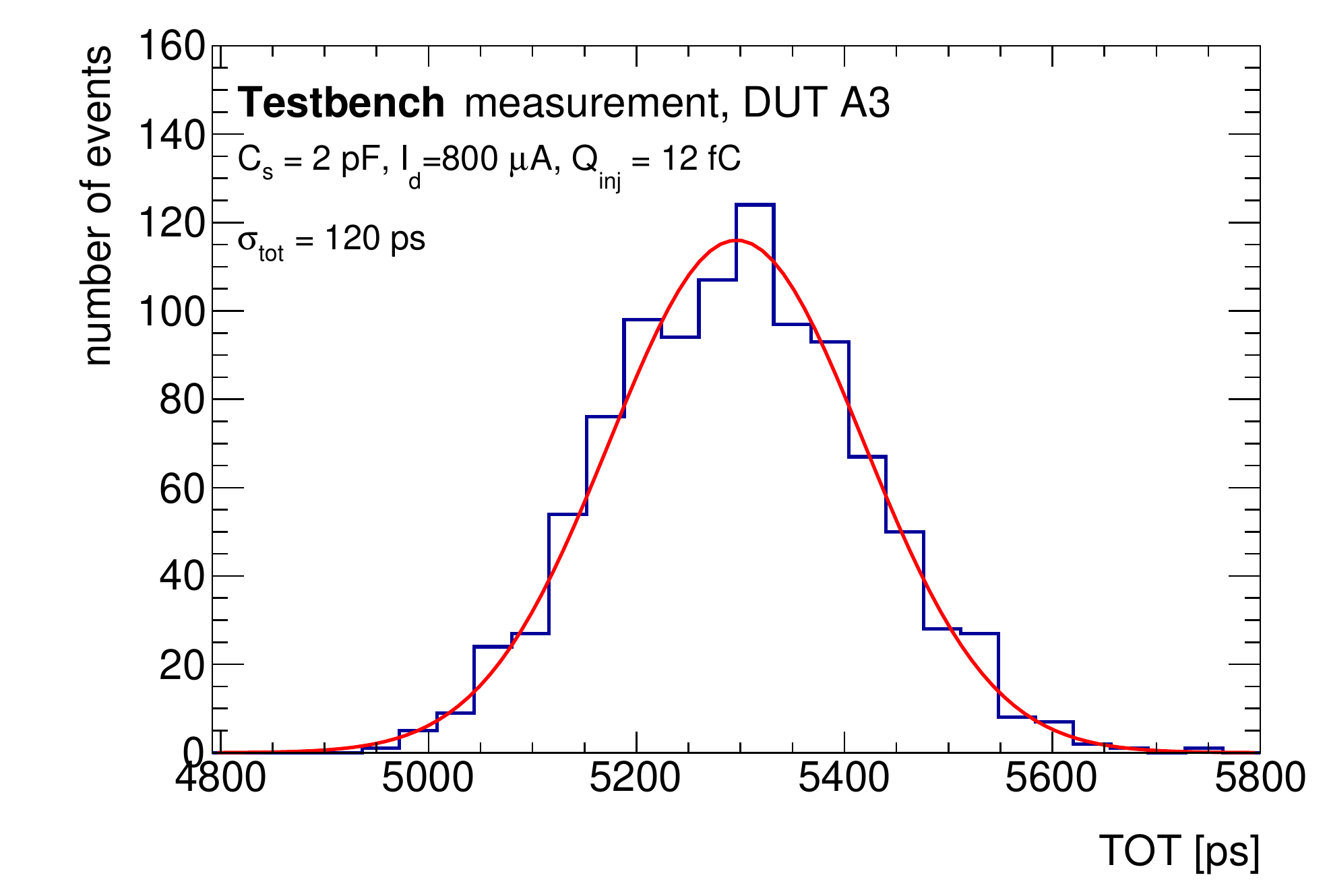}\label{fig:tot1D_12fC}}
\hspace*{0.2cm}\subfloat[]{\includegraphics[width=0.5\textwidth]{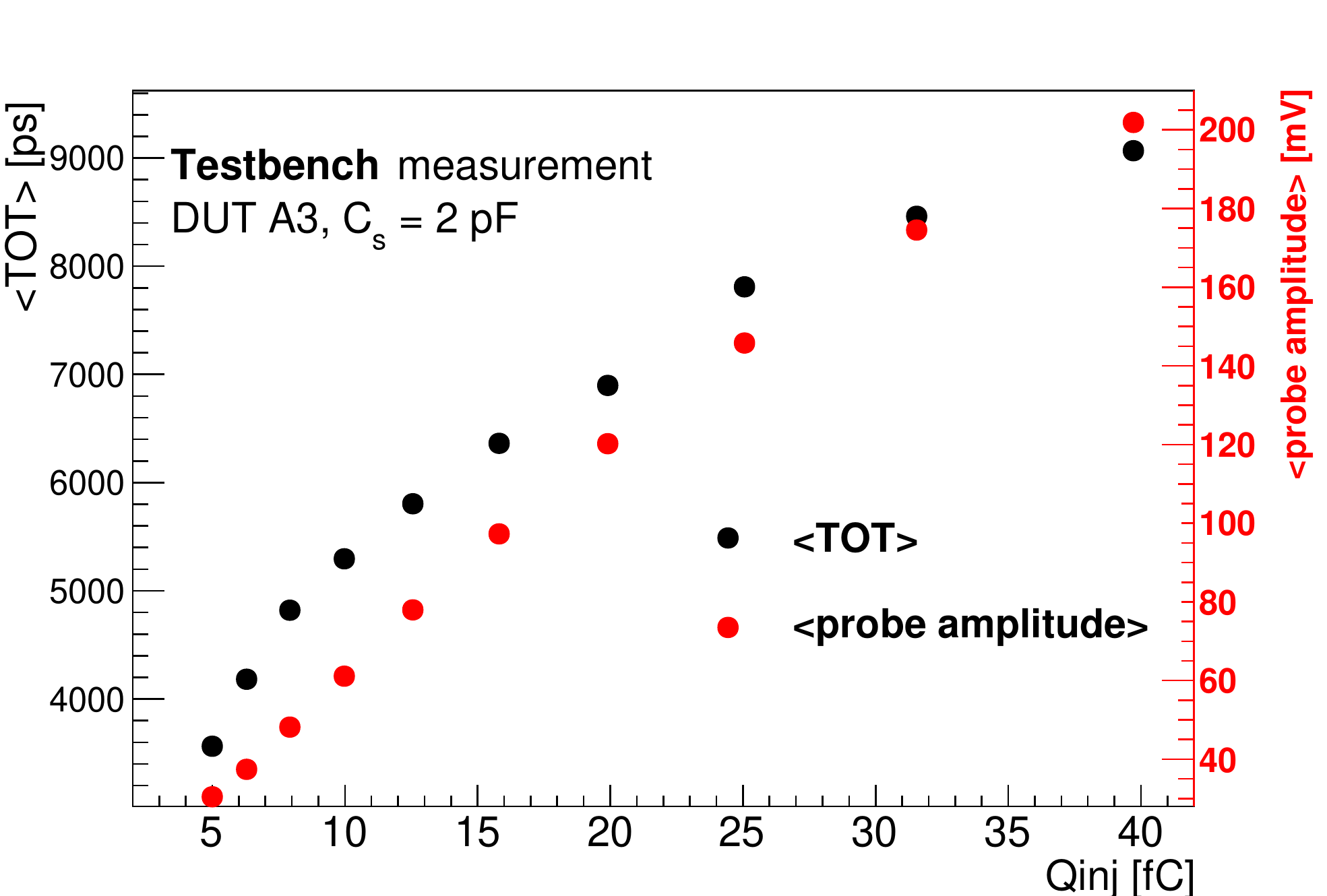}\label{fig:totAmp_vs_q}}

\subfloat[]{\includegraphics[width=0.5\textwidth]{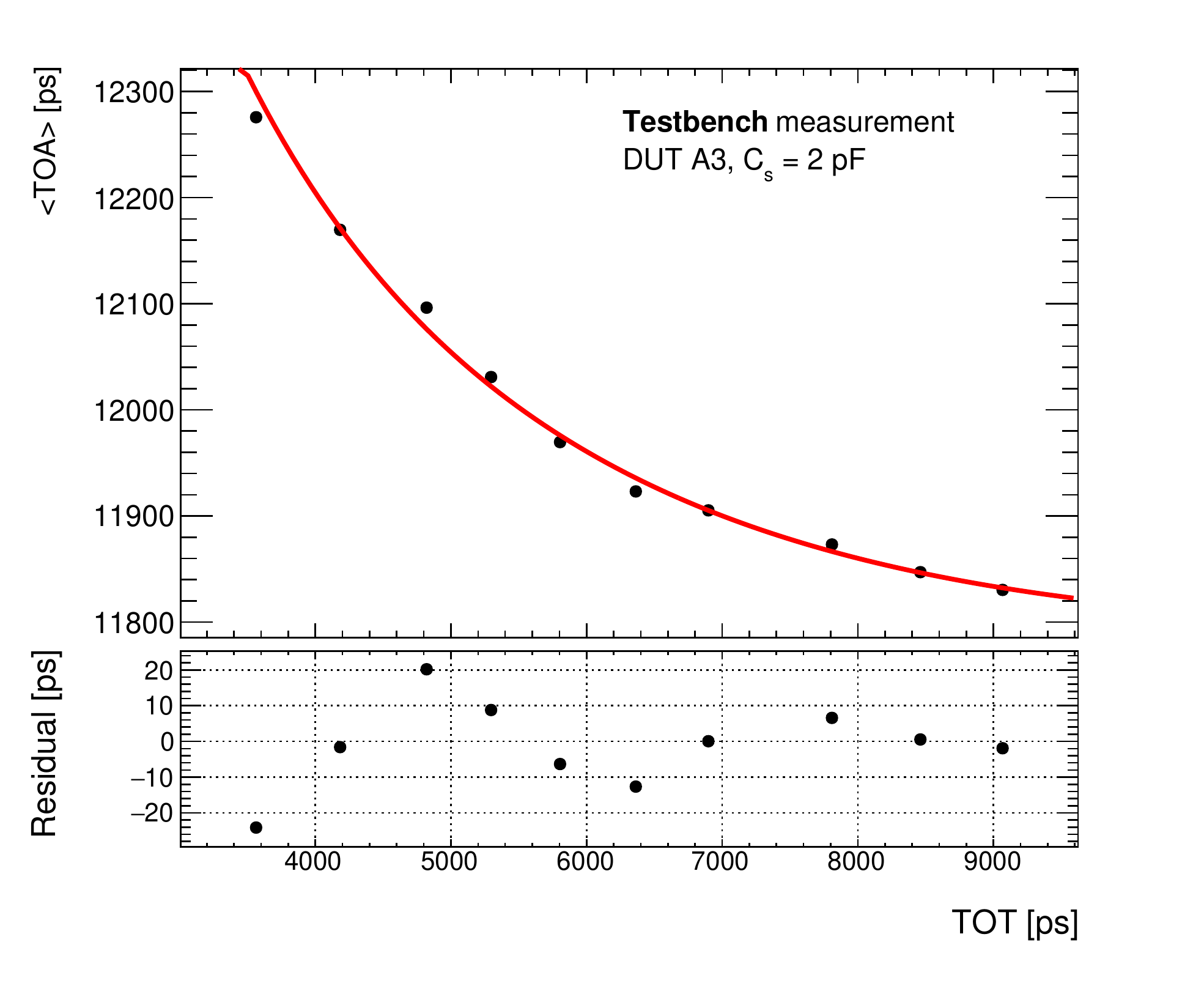}\label{fig:toavstot}}
\hspace*{0.2cm}
\subfloat[]{\includegraphics[width=0.5\textwidth]{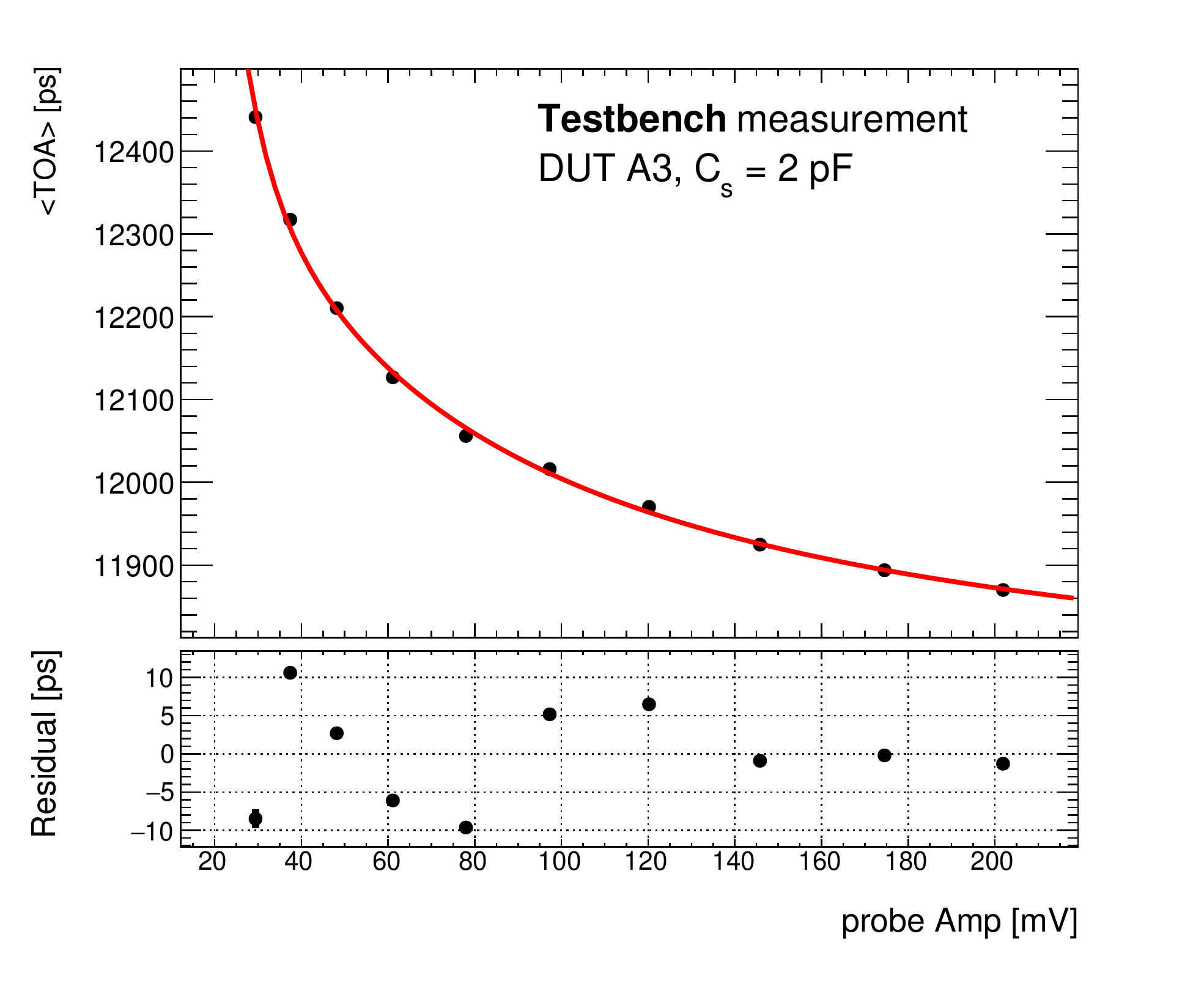}\label{fig:toavsProbe}}

\caption{ (a) Distribution of the discriminator TOT for Q$_{inj}$ = 12 fC. (b) Average TOT (in black) and probe amplitude (in red) as a function of the injected charge. (c) Average time of arrival as a function of the average time over threshold for various injected charges. (d) Average time of arrival as a function of the average probe amplitude for various injected charges. The fit used for the time walk correction is superimposed (red line) in both cases, while the bottom plot in both figures shows the residual of the average after the time walk correction. A 2 pF capacitance has been soldered to emulate the sensor capacitance.}

\label{fig:timewalk}

\end{figure}

\end{subsubsection}

\end{subsection}

\section{Test bench module performance with ALTIROC0 }
\label{sec:performance_lgad}

The sequence of measurements shown in chapter~\ref{sec:performace_asic} has been repeated with 
the ASIC bump bonded to the sensor, seen as a capacitance C$_d$. The tested sensors were always operated at a bias voltage of V$_{bias}$ = -90 V. This operating point was chosen to ensure their full depletion. The leakage current of the modules at this bias voltage was measured to be of the order of 10$^{-2}$ \SI{}{\micro\ampere}, a value that has a negligible impact on the overall performance of the devices.

\subsection{Jitter}

\begin{figure}[H]
\centering

\includegraphics[width=0.5\textwidth]{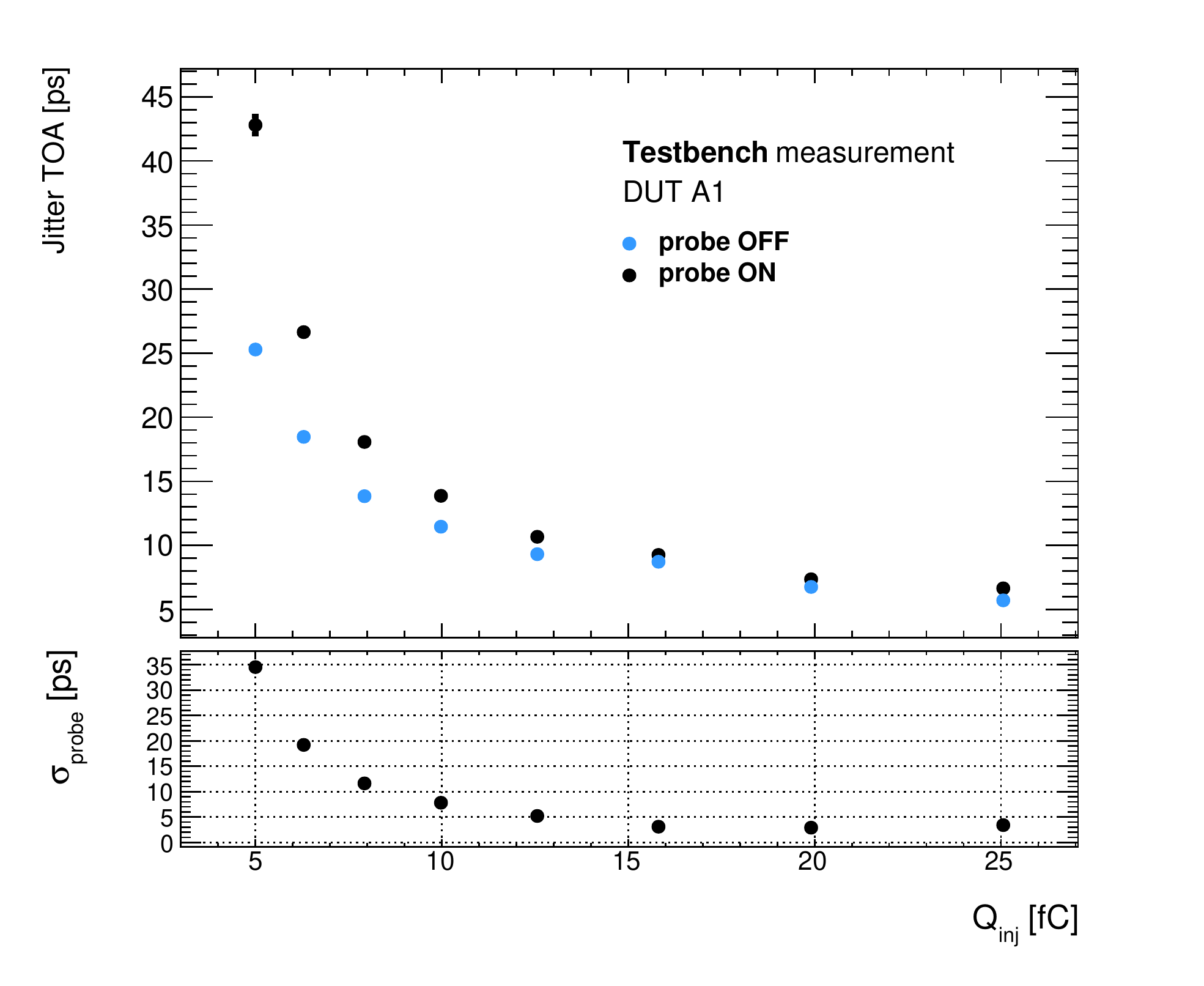}

\caption{
  (up) Discriminator jitter as a function of the injected charge, after subtraction of the generator time resolution. The two sets of measurements correspond to a configuration where the preamplifier probe is either active or inactive. (down) Probe contribution to the time resolution, defined as the quadratic difference of the TOA jitter between having the probe active or inactive. The activation of the preamplifier probe naturally degrades the performance.  }
  \label{fig:jitterWithSensor}

\end{figure}

The TOA jitter as a function of the injected charge is shown in Figure \ref{fig:jitterWithSensor} for two configurations; one with the preamplifier probe turned off and the other with the probe activated.  In both cases, a constant threshold equivalent to 2.5 fC is used. When the probe is not activated, it is found that, for 5 fC, the measured jitter is 25 ps, while for 10 fC it is approximately 13 ps. These results are consistent with the ones presented in section \ref{subsubsec:jitter_asic} where the ASIC was without a sensor and with a soldered capacitance of 2 pF. The activation of the probe naturally degrades the discriminator performance due to an increase of the preamplifier rise-time. The probe contribution to the time resolution, $\sigma_{probe}$, defined as the quadratic difference of the TOA jitter between having or not the probe active, can be extracted from calibration, as seen in the bottom pad of Figure \ref{fig:jitterWithSensor}. This contribution shows a strong dependence on the injected charge. It is found to be 8 ps for Q$_{inj}$ = 10 fC and reaches a negligible value of 4 ps for Q$_{inj} >$ 15 fC.

\subsection{Measurements at cold temperature}
\begin{figure}[H]
\centering
\subfloat[]{\includegraphics[width=0.5\textwidth]{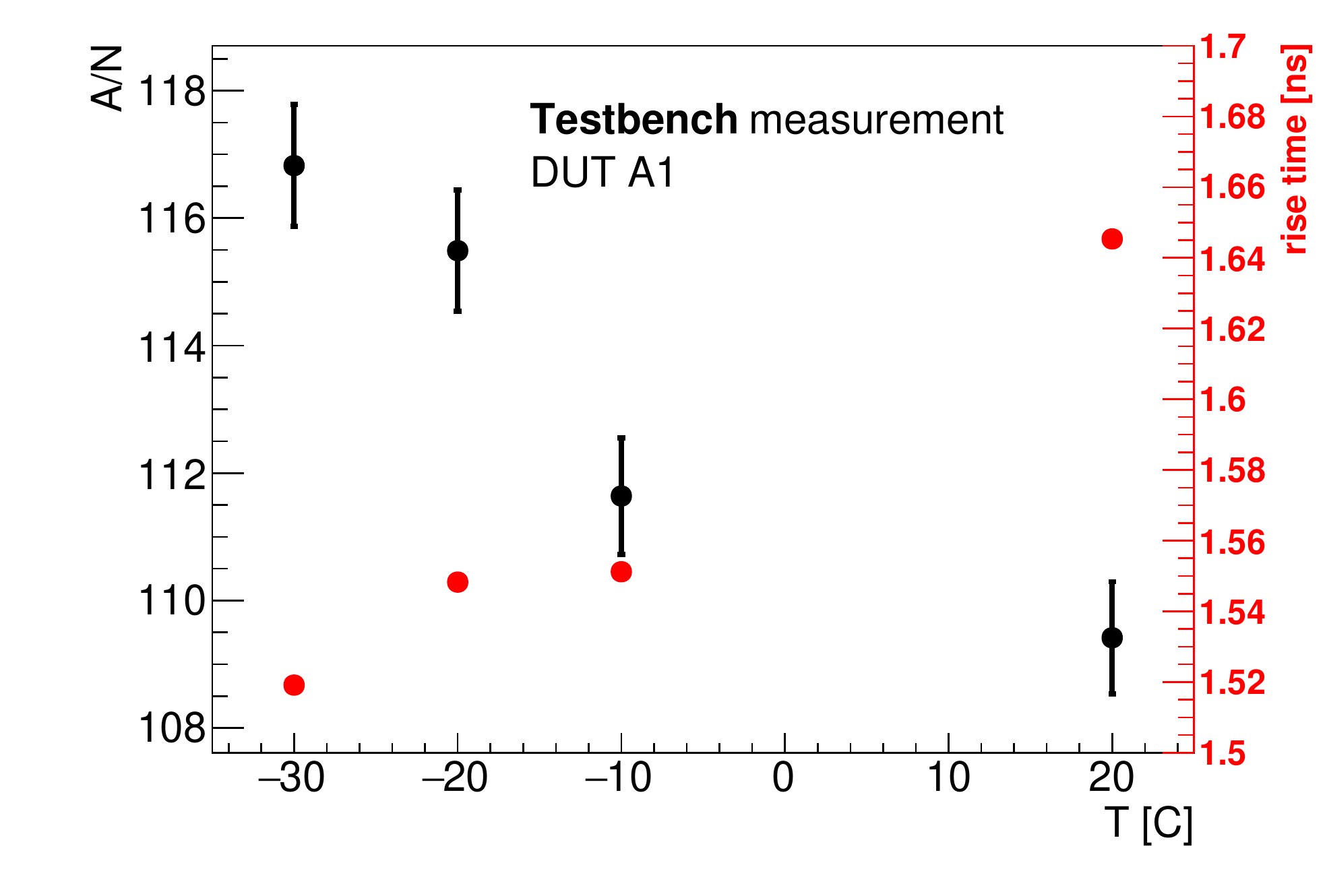}\label{fig:SoverN_temp}}
\hspace*{0.2cm}
\subfloat[]{\includegraphics[width=0.5\textwidth]{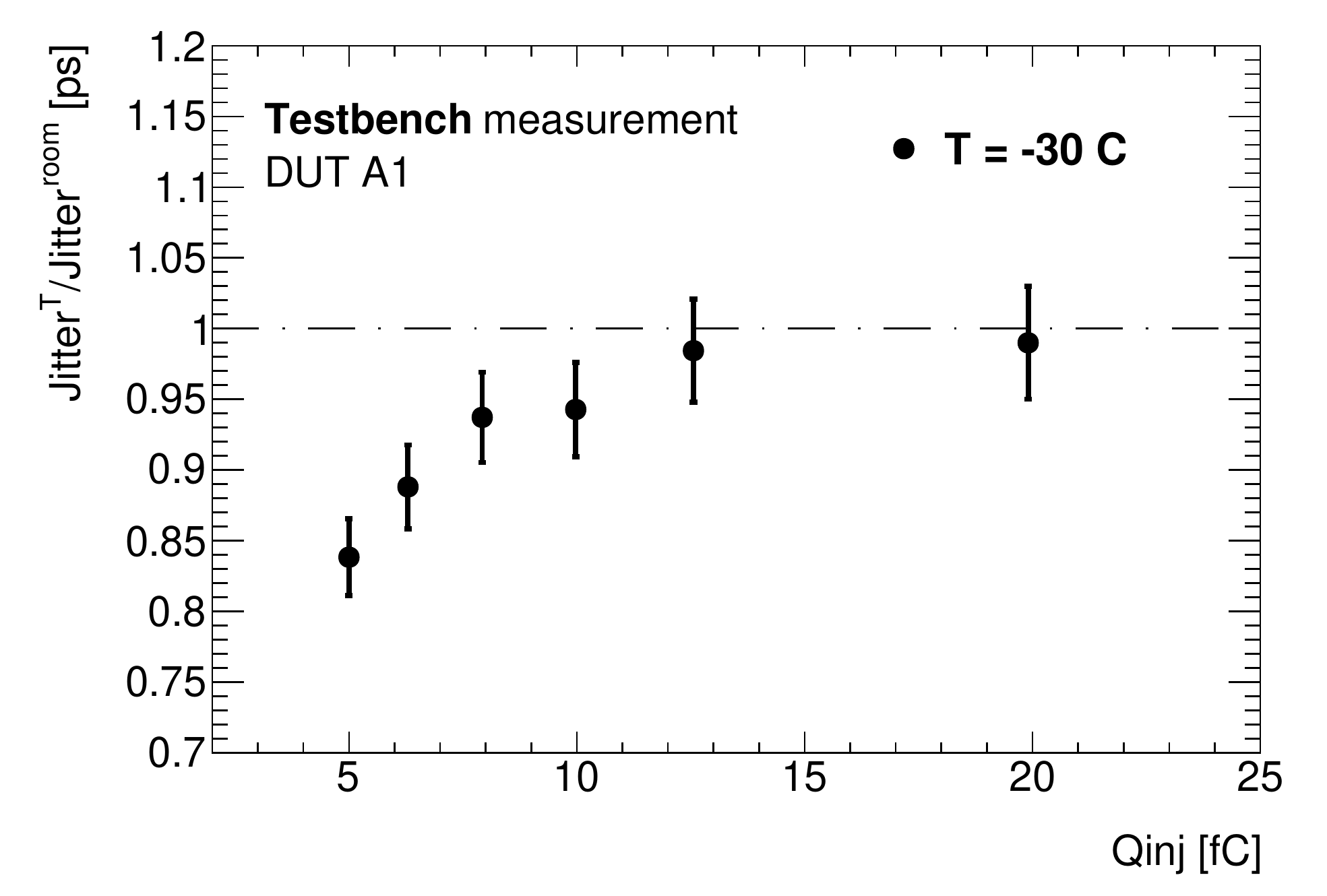}\label{fig:jitter_temp}}

\caption{ (a) Amplitude over noise ratio (in black) and rise-time (in red) as a function of the temperature for an injected charge of 10 fC. (b) Ratio of the TOA jitter at T = -30 $^o$C to the jitter at room temperature as a function of the injected charge.}

\label{fig:coldMeasurements}
\end{figure}

Within the HGTD, the ASIC is expected to operate down to -30$^o$ C in order to mitigate the increase of the sensor leakage current with irradiation. For this reason, the performance of ALTIROC0 was studied using a climate chamber, constantly supplied with dry air to avoid condensation. The results are shown in Figure \ref{fig:coldMeasurements}.
The signal amplitude over noise ratio at the preamplifier output can be estimated from the probe. As shown in Figure \ref{fig:SoverN_temp}, there is a 7\% increase in the A/N between 20 and -30 $^o$C. In parallel, the probe rise-time, defined as the difference between the time when the pulse is at the 10\% and 90\% of its maximum amplitude, decreases with the temperature by the same order of magnitude as the increase in A/N.\footnote{This behaviour of the rise-time could originate from the probe shaper and not the preamplifier itself.} 

Figure \ref{fig:jitter_temp} shows the TOA jitter as a function of the injected charge. As expected, the jitter improves when moving to lower temperatures. The effect of the temperature on the jitter arises from the temperature dependence of the noise and of the transistor transconductance $g_{m1}$, the latter of which is directly proportional to V$out_{pa}$. Indeed, by combining equations \ref{eq:transconductance} and \ref{eq:noise_density}, it follows that the N/A, and therefore, the jitter is proportional to the square of the temperature. This effect is more prominent for low values of the injected charge, whereas it becomes less pronounced for values above 10 fC, due to the saturation of the preamplifier. An overall reduction of the jitter of the order of 6\% is observed for a Q$_{inj}$ = 10 fC at the lowest temperature point. While this reduction follows the expected trend, it is less pronounced than the combined effect expected from the simultaneous increase (decrease) of the preamplifier signal amplitude over noise (rise-time) with temperature. For the same injected charge, the latter is of the order of 14\%. This behaviour is not fully understood.

\subsection{Time walk correction}
 For the modules which include an LGAD sensor, a different way to apply the time walk correction was developed. The problem that led to this new approach along with the performance of the modified method for the time walk correction are presented below.

\subsubsection{Time-Over-Threshold problem}\label{subsec:totProblem}

\begin{figure}[H]
\centering
\subfloat[]{\includegraphics[width=0.5\textwidth]{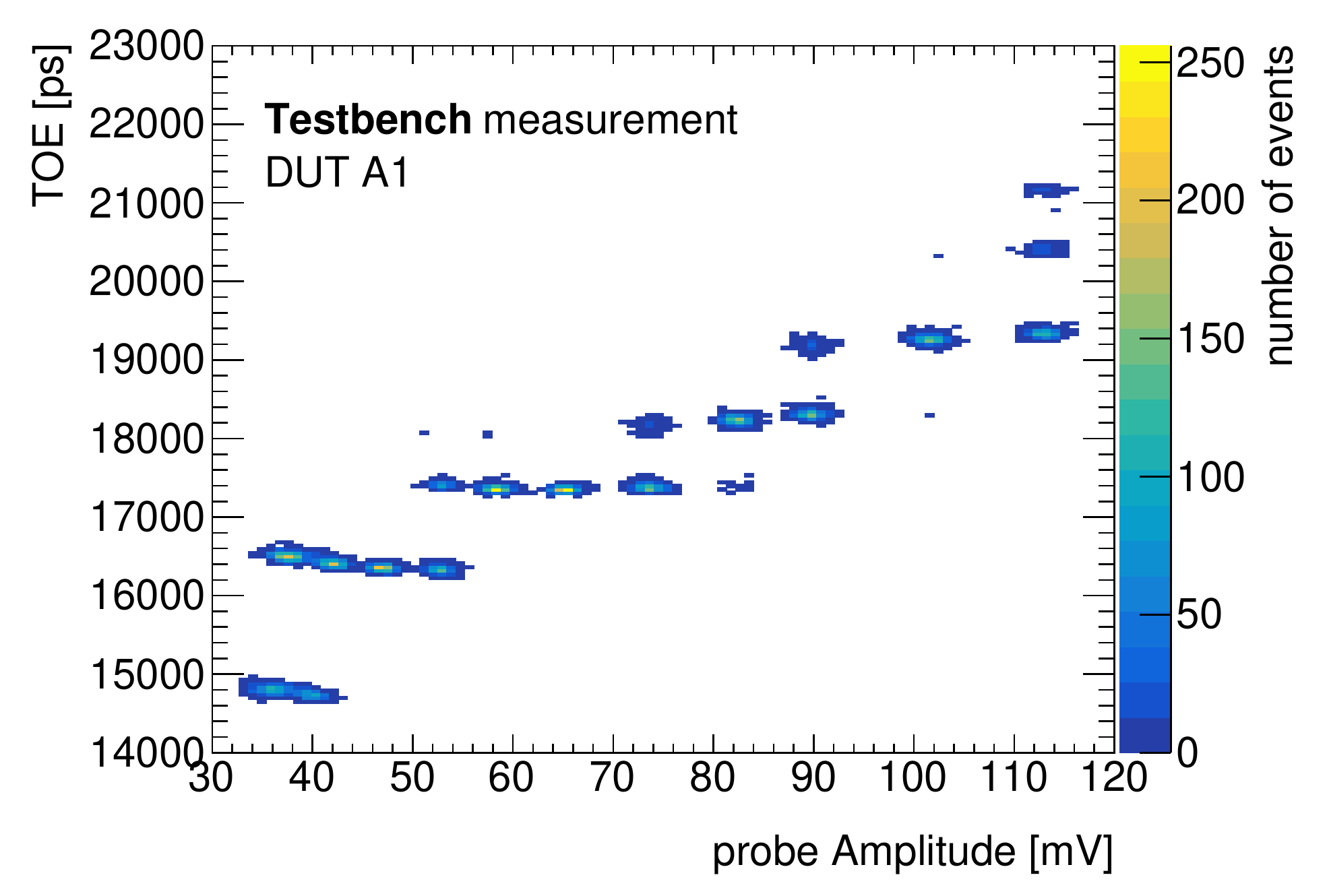}\label{fig:totProblem_b23}}
\hspace*{0.2cm}
\subfloat[]{\includegraphics[width=0.5\textwidth]{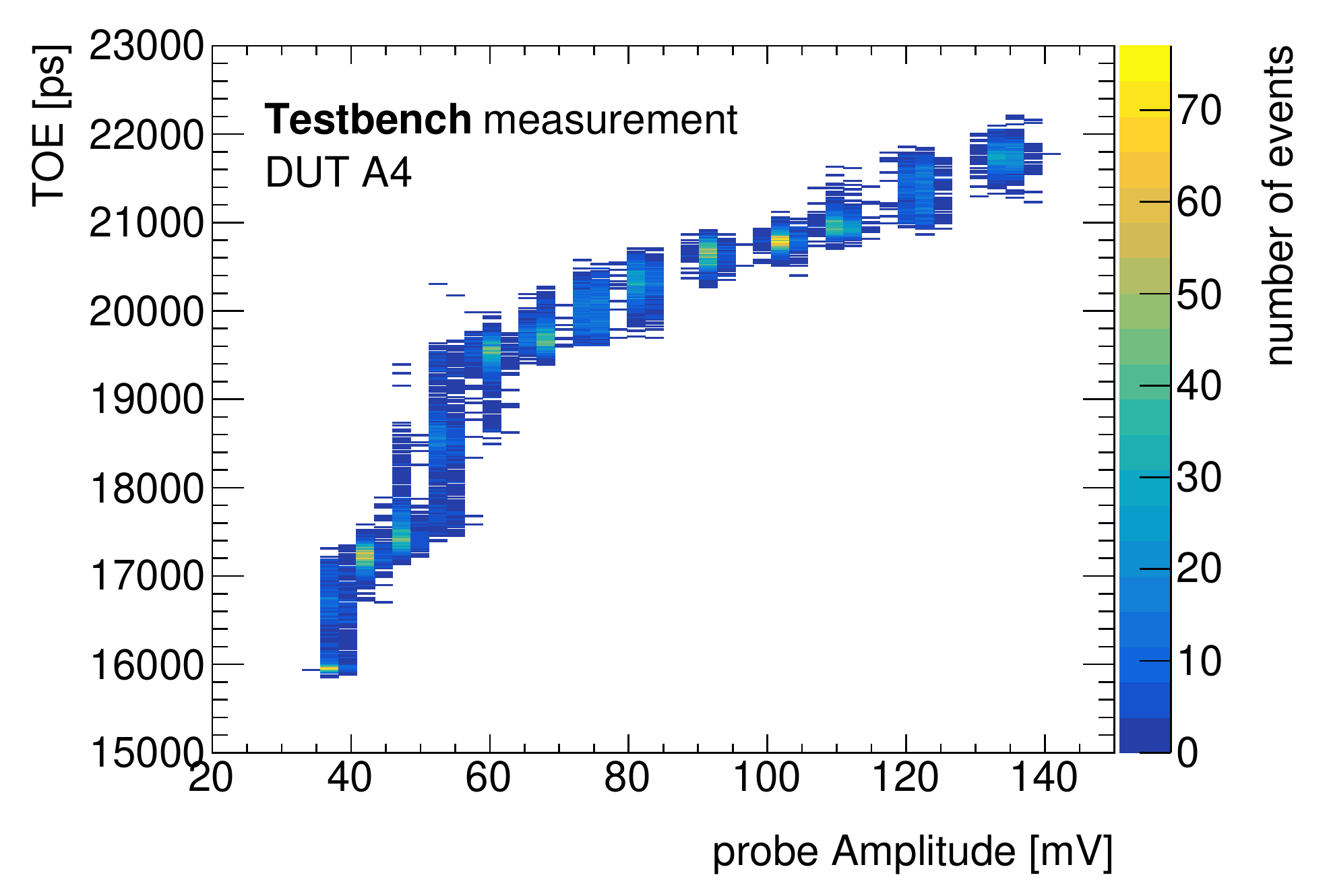}\label{fig:totProblem_b32}}

\caption{TOE as a function of the probe amplitude for various injected charges, for (a) a standard HV-connection board and (b) an L-shaped HV pad board. Both boards are equipped with an ASIC bump-bonded to an unirradiated 2x2 sensor array. }

\label{fig:totProblem}
\end{figure}

It was observed that the discriminator TOT does not scale proportionally with the probe amplitude when a sensor is bump bonded to the ASIC. Moreover, afterpulses were observed on the discriminator falling edge.
Figure \ref{fig:totProblem_b23} shows how, when using a calibration pulse with a charge ranging from 5-20 fC, the time of end (TOE) of the pulse presents a discrete behaviour with respect to the probe amplitude, while the relation should be continuous. Two possible reasons for this problem have been theorised; an inductance caused by the length of the pad-sensor HV connection or a coupling of the discriminator output to the PCB. 

\begin{figure}[H]
\centering
\includegraphics[width=0.5\textwidth]{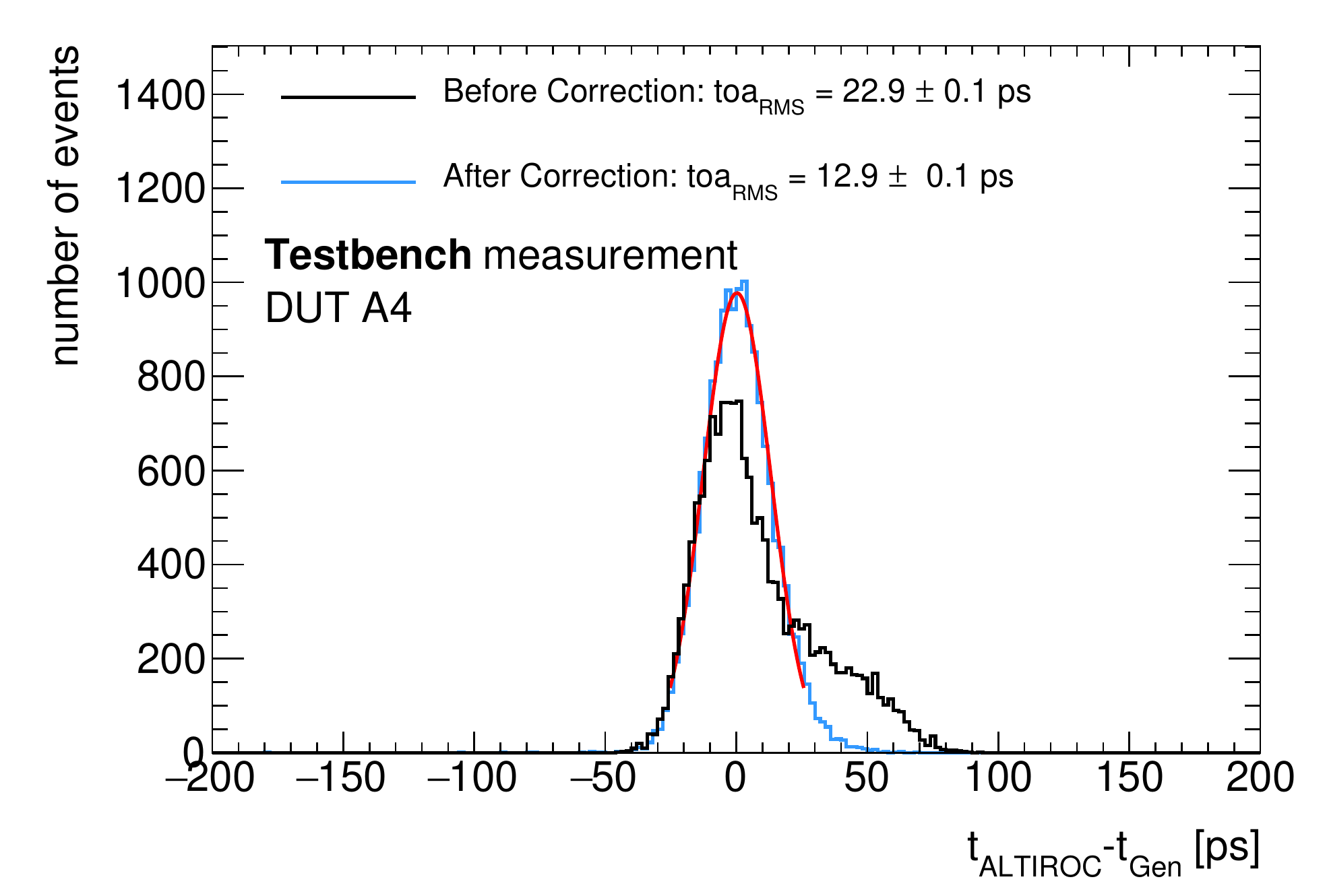}

\caption{TOA distribution for a charge between 12 and 20 fC before and after time walk correction for an L-shaped HV pad board. The board is equipped with an ASIC bump-bonded to an unirradiated 2x2 sensor array. The time walk has been corrected using the discriminator TOT. A Gaussian fit (red line) is applied to the corrected distribution. }

\label{fig:residual_b33}
\end{figure}

In order to investigate the former, a new board with a larger L-shaped HV connection pad was manufactured. This particular shape of the HV pad allows for many wire bonds to be attached far from each other in order to reduce any possible inductance. It can be seen in Figure \ref{fig:totProblem_b32} that the issue is still present in the modified board for Q$_{inj} <$ 10 fC (corresponding to probe amplitude < 60 mV). However, it is reduced for higher charges. Figure \ref{fig:residual_b33} shows the distribution of the TOA in the reduced charge range between 12 - 20 fC, before and after applying a time walk correction using the TOT. 
The time walk correction results in a 40\% improvement of the TOA RMS, which is found to be 13 ps after subtracting the generator resolution.
For the second version of the chip, ALTIROC1, an L-shaped pad has been implemented.

\subsubsection{Correction using the preamplifier probe}
\begin{figure}[H]
\centering
\includegraphics[width=0.5\textwidth]{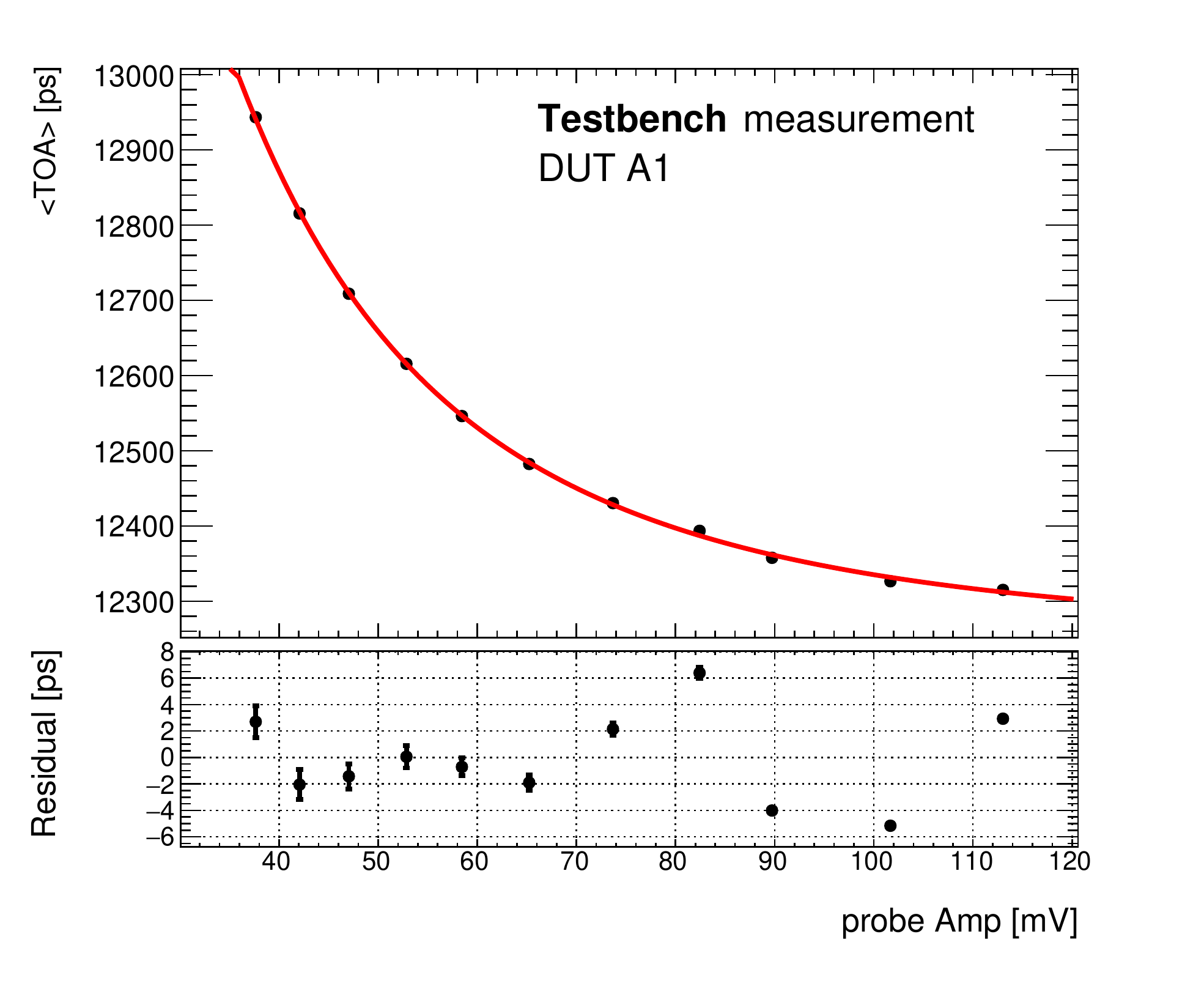}
\caption{(up) Average time of arrival as a function of the average probe amplitude. The fit used for the time walk correction is superimposed (red line). (down) Residual of the average after the time walk correction.}
\label{fig:timewalkWithSensor}

\end{figure}

Due to the discreteness problem in the falling time, the discriminator TOT was not chosen as the default method for the time walk correction
in ALTIROC0 boards with the standard HV pads.  Instead, the time walk was corrected using the probe amplitude. As shown in Figure \ref{fig:timewalkWithSensor}, the correction in a charge range of 5 to 20 fC results in a residual with a peak-to-peak variation of 12 ps. This outcome is compatible with measurements of the ASIC alone and within the requirements of the HGTD. 
The time walk correction using the probe amplitude was used for the testbeam measurements, since only boards with the standard HV pads were available at that time.

\section{Test Beam module performance with ALTIROC0 }
\label{sec:performance_testbeam}

Two  modules were exposed to 120 GeV pions 
at the H6B beamline at the CERN-SPS North Area during one week in October 2018. This section presents the 
results collected during this data taking period. 

\begin{subsection}{Testbeam Setup}

The pulses of 2$\times$2 LGAD sensor arrays mounted on up to 2 ALTIROC0 boards were sampled by two Agilent Infiniium DSA91204A oscilloscopes with a 40 GSample/s sampling rate and a bandwidth of 12 GHz. For an accurate timing reference, two fast Cherenkov trigger counters were used. Each one consists of a Cherenkov-light emitting Quartz bar of 3$\times$3 mm$^2$ area transverse to the beam and 10 mm length along the beam, coupled to a Silicon Photomultiplier (SiPM). The time resolution of these devices was measured to be about 40ps. 

A EUDET-type beam telescope~\cite{telescope}  based on MIMOSA pixel planes with a track position precision of few micrometres  was also included in the data taking, allowing for position-dependent measurements. The trigger was provided by the coincidence of signals on a scintillator and a special 3D FE-I4 plane \cite{FEI4}. More details on the tracking and trigger configuration can be found in~\cite{TestBeamPaper}, where a similar setup was used.

Custom-made support structures provided mechanical stability of the ALTIROC and SiPM boards. The ALTIROC DUTs were mounted on a base
plate integrated in the EUDET telescope. A separate base plate was used for the positioning of the SiPM devices, while a styrofoam box ensured their light-tightness. 
Remotely controllable stage motors allowed for movement in the horizontal and vertical directions perpendicular to the beam direction with micrometre precision of both base plates. This allowed for a precise positioning of the sensors at the centre of the beam and alignment of the DUTs to the SiPMs. 
\end{subsection}

\begin{subsection}{Results}
For all the results presented hereafter, both modules were operated at a voltage of V$_{bias}$ = - 120 V, to ensure the depletion of the sensor and a high gain. The leakage current was continuously monitored and was always found to be of the order of 10$^{-2}$ \SI{}{\micro\ampere} for both sensors.

\begin{subsubsection}{Pulse properties}
The probe amplitude of one channel of DUT A1 and A2 is shown in Figure \ref{fig:amp_B23vsB24} for the operating point of V$_{bias}$ = -120 V. It can be seen that the two DUTs behave similarly with A2 showing slightly larger amplitude.  
The deposited charge in the sensor in testbeam is calculated from the integral of the preamplifier probe pulse. Calibration measurements with the picosecond pulse generator, in which the injected charge is known with high accuracy, are used to determine the relation of the probe integral to the injected charge. This relation is then used to extract the equivalent deposited charge in the sensor in testbeam. The resulting distributions for a channel of DUT A1 and A2 are shown in Figure \ref{fig:charge_B23vsB24}. Both distributions are fitted with a Landau function convoluted with a Gaussian function to extract the most probable value.
It was found that, for V$_{bias}$ = -120 V, the most probable injected charge in testbeam was Q$_{inj}$ = 18 fC for DUT A1 and Q$_{inj}$ = 24 fC for DUT A2.\footnote{These values correspond to a sensor gain of 35 and 47, respectively.}  While, in both DUTs the charge is higher than the planned benchmark point for the HGTD, it should be noted that the goal of the measurements presented here was the initial characterization of ALTIROC+LGAD un-irradiated modules. The study of the module performance at the lowest limit of the ALTIROC dynamic range is planned for future campaigns.

\begin{figure}[H]
\centering

\subfloat[]{\includegraphics[width=0.5\textwidth]{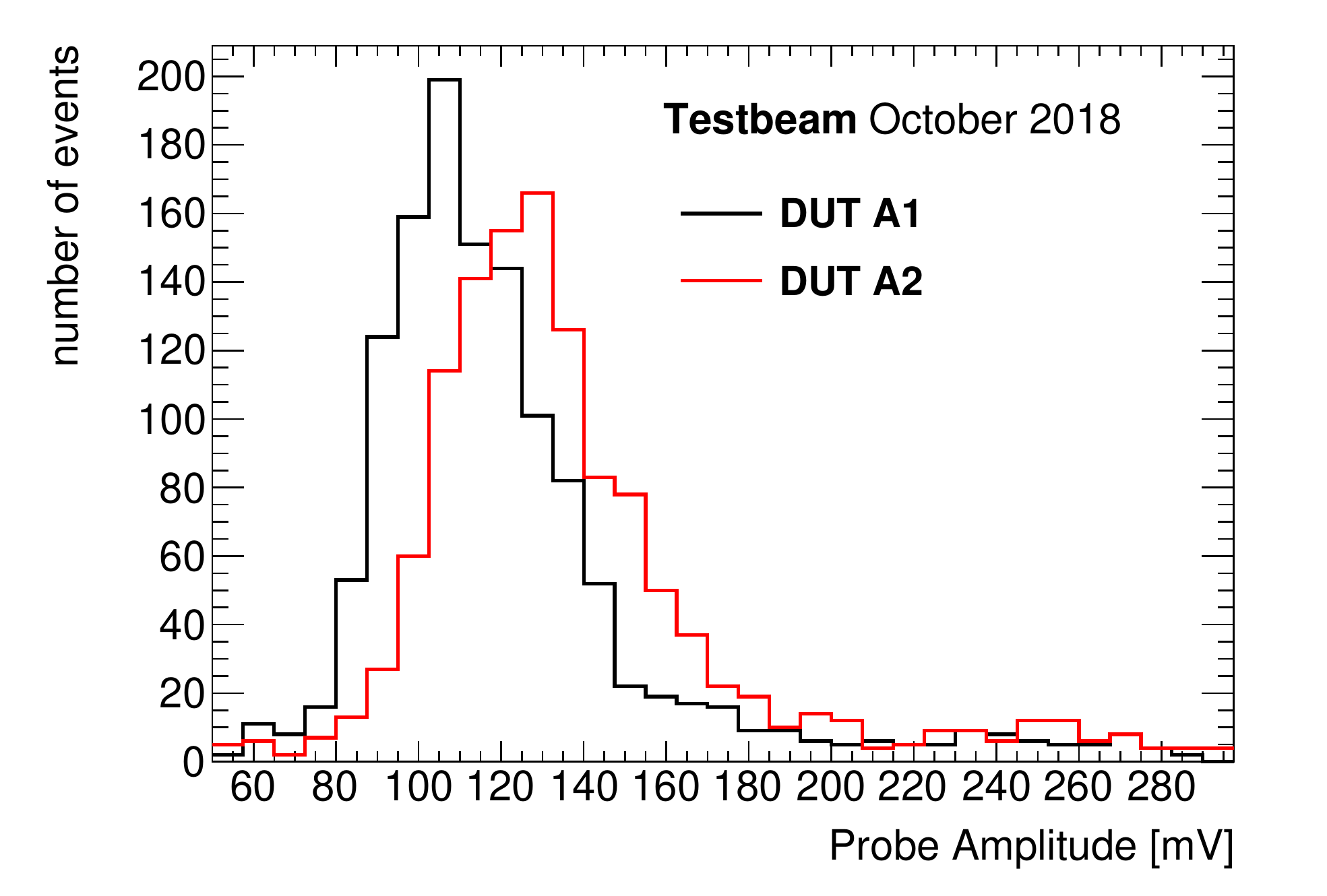}\label{fig:amp_B23vsB24}}
\hspace*{0.2cm}
\subfloat[]{\includegraphics[width=0.5\textwidth]{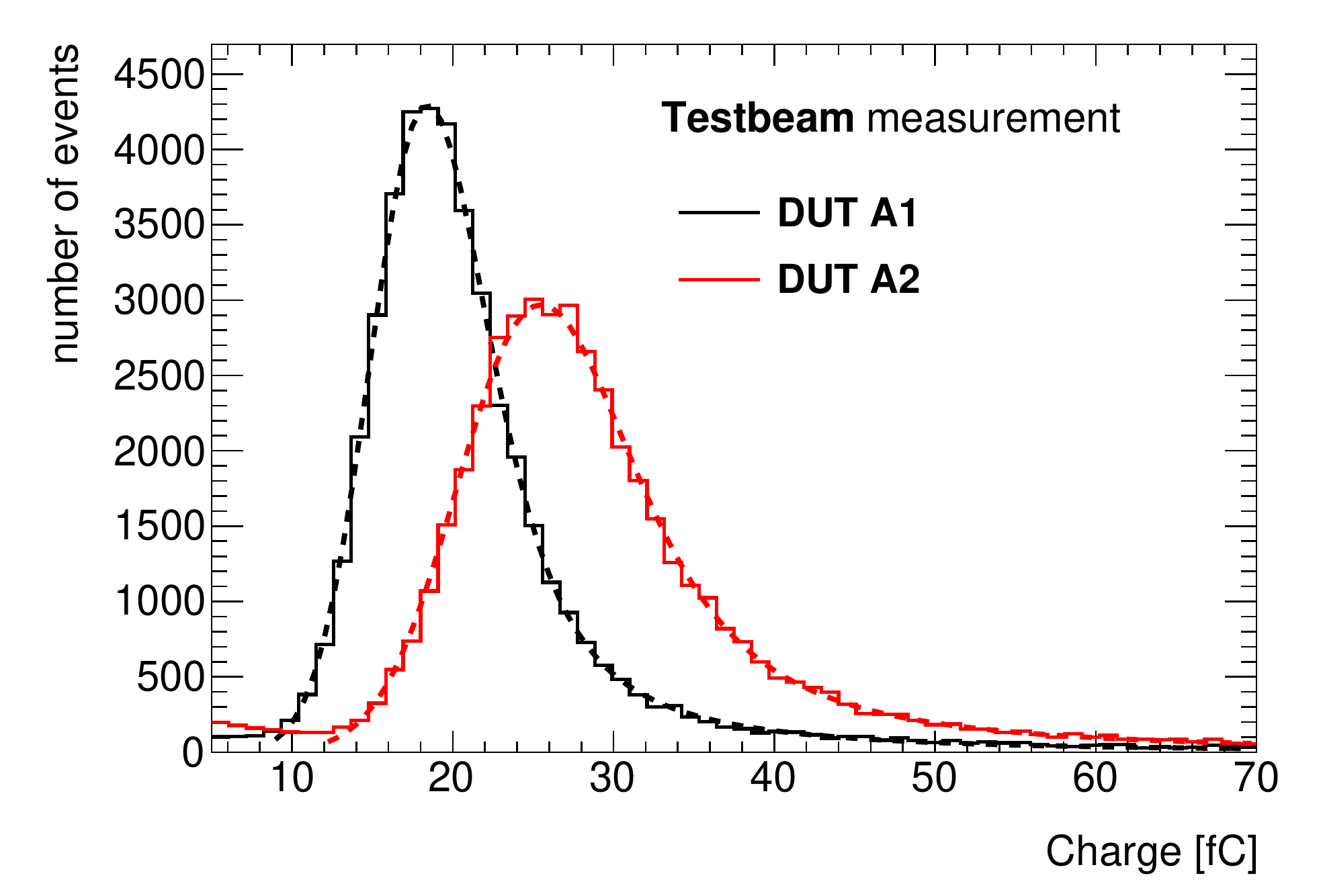}\label{fig:charge_B23vsB24}}

\caption{Distribution of (a) the probe amplitude and (b) the deposited charge for the same ASIC channel of DUT A1 and A2 for a bias voltage of -120 V. A Landau function convoluted with a Gaussian function is fitted in each charge distribution to extract the most probable deposited charge. The functions are displayed as dashed lines of the same colour as the fitted distributions.}
\end{figure}

\end{subsubsection}

\begin{subsubsection}{Time measurement performance}

The time resolution of the DUT is estimated from the time difference between the time of arrival (TOA) of the DUT and the SiPM. The TOA is defined as the time at half of the maximal amplitude of the considered signal. The DUT resolution is the convolution of the jitter of the electronics, the Landau fluctuations of the sensor and the time walk effect. This last contribution can be corrected. Due to the discrete behaviour of the discriminator falling edge that was discussed in section \ref{sec:performance_lgad}, the probe amplitude is used to correct for the time walk effect, of about 200 ps, as shown in Figure \ref{fig:toa_vs_amp_B24}. The probe contribution to the time resolution is negligible for Q$_{inj} \geq$  18 fC, as demonstrated in Figure \ref{fig:jitterWithSensor}. 

After correction of the time walk effect, the time difference is also shown in Figure \ref{fig:propagandaPlot}, where a Gaussian fit is applied. The expected time resolution of the SiPM (40 ps) is quadratically subtracted. The time walk correction brings on an improvement by a factor of 30\%.
The time resolution of each channel of the two DUTs after correction is summarised in Table \ref{tab:timeReso}. It should be  noted that the DUT A2 provides systematically a better resolution. This can be explained by the larger charge of A2, as shown in Figure \ref{fig:charge_B23vsB24}. With calibration signals, this amplitude difference is not observed, therefore, it has been traced back to a different gain of the LGAD sensors. The performance is better than 40 ps for all channels of the A2 DUT, with a best achieved time resolution of 34.7 ps after time walk correction. 

\begin{figure}[H]
\centering
\subfloat[]{\includegraphics[width=0.5\textwidth]{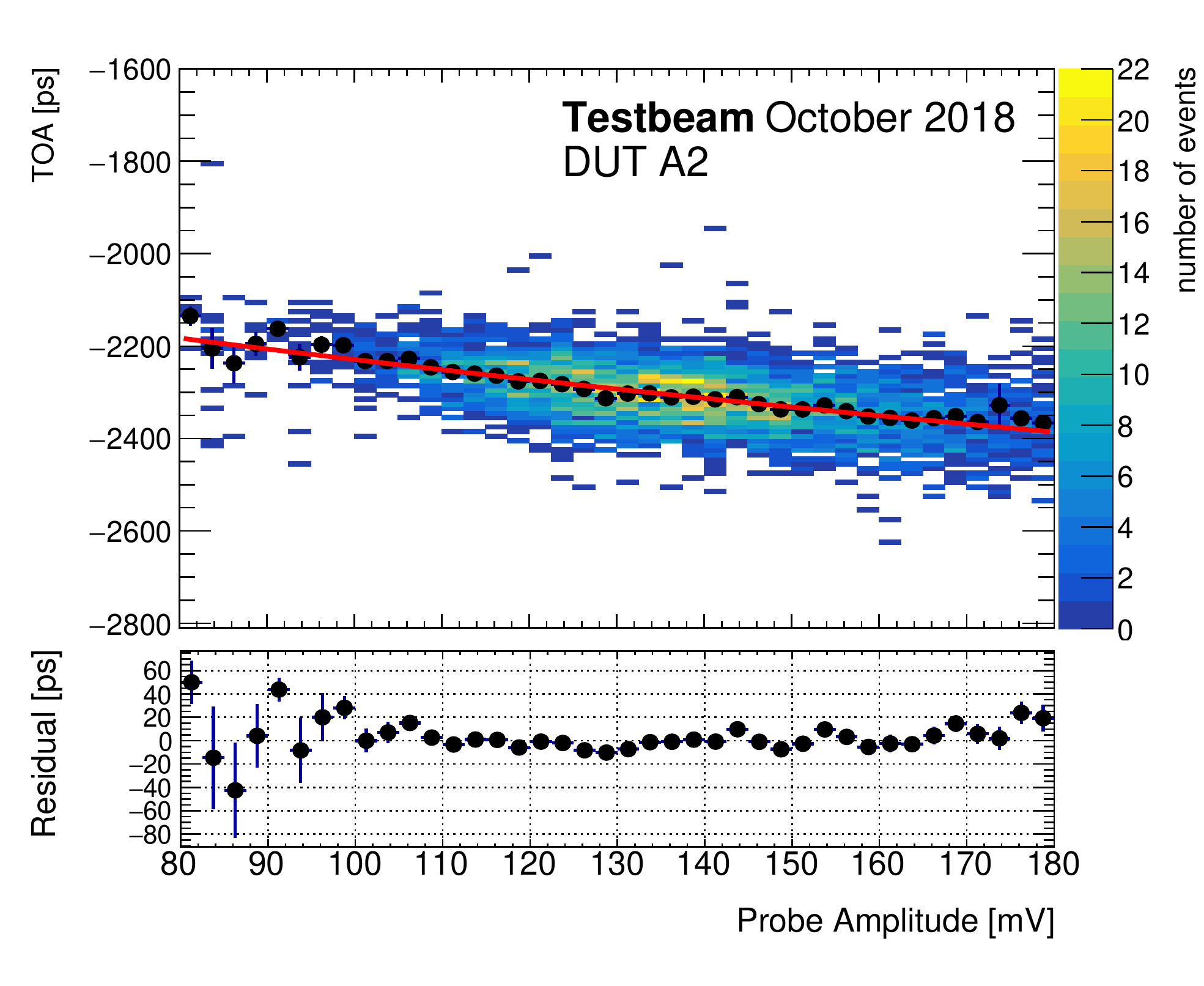}\label{fig:toa_vs_amp_B24}}
\hspace*{0.2cm}
\subfloat[]{\includegraphics[width=0.5\textwidth]{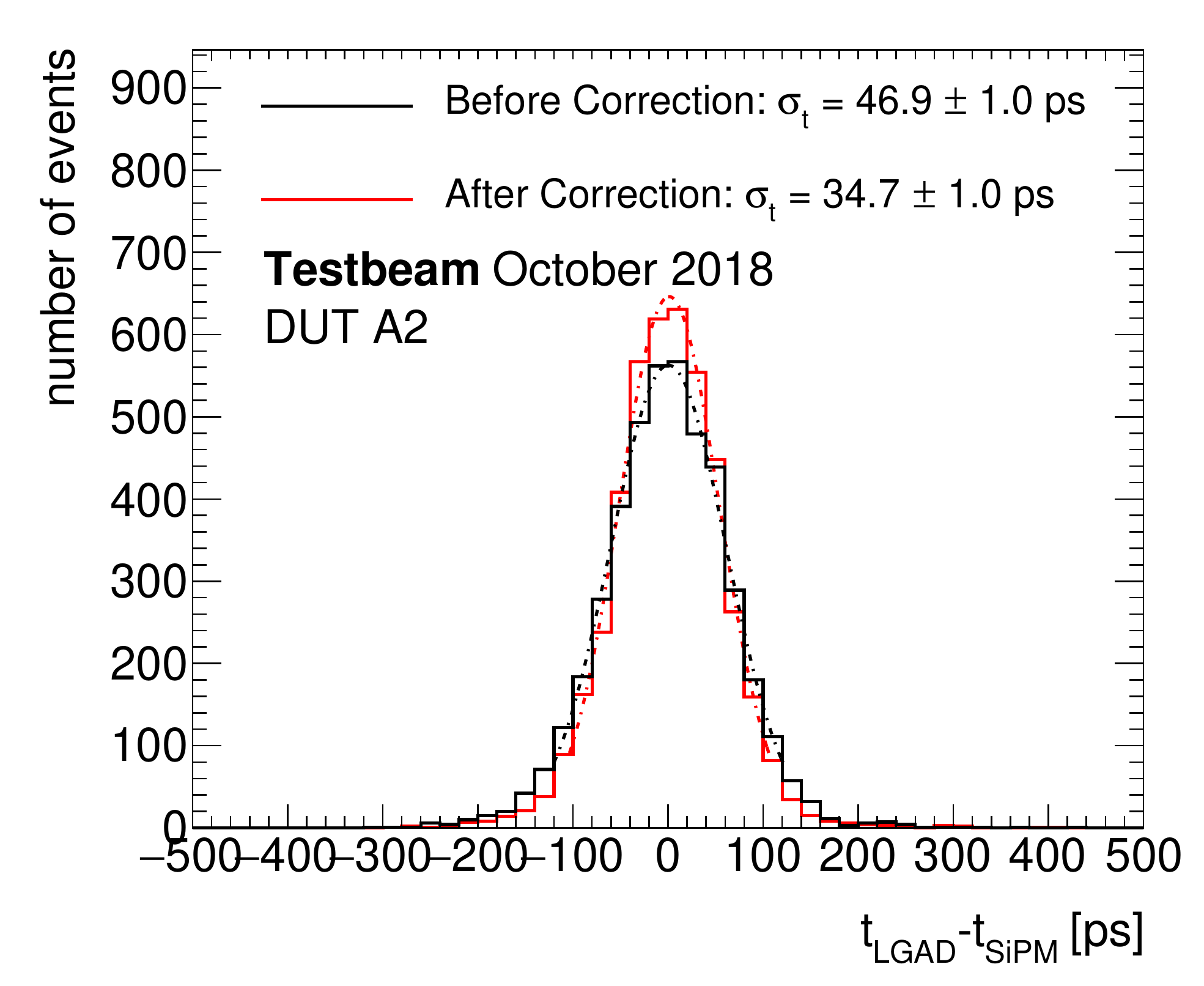}\label{fig:propagandaPlot}}

\caption{(a) (up) Time of arrival difference for a channel of an ALTIROC0-LGAD bare module as a function of the preamplifier probe amplitude. The profile of the 2D distribution (black points) and a polynomial fit (red line) are superimposed. The fit is used to correct for the time walk effect. (down) Residual of the average time of arrival difference after the time walk correction as a function of the preamplifier probe amplitude. (b) Distribution of the time of arrival difference for a channel of the ALTIROC0-LGAD bare module before 
and after time walk correction. A quartz+SiPM is used as a time reference. A discriminator threshold of 4.5 fC was used for this measurement. }

\end{figure}

This value was compared to a calibration run reproducing as close as possible the testbeam conditions; a jitter of 7 ps was found in this case for the testbeam-equivalent injected charge of Q$_{inj}$=24 fC. Taking into account the sensor Landau contribution, which is known to be around 25 ps for un-irradiated LGADs \cite{nicolo} \cite{TestBeamPaper}, as well as the deterioration of the jitter (by a factor of 1.65) due to the longer duration of the LGAD signal compared to the calibration pulse, results in a performance of $\sim$ 27 ps. Finally, adding in quadrature  a pessimistic peak-to-peak value of the time walk correction residual, already extracted from Figure \ref{fig:timewalkWithSensor}, gives a time resolution of $\sim$ 30 ps. This result is compatible with the best achieved testbeam performance.

The time resolution was also measured as a function of the discriminator threshold, as shown in 
Figure \ref{fig:reso_DAC_B24}. The threshold was varied from 153 to 173 mV, corresponding to a Q$_{inj}$ ranging approximately from 2 to 8 fC; a small increase is observed for larger threshold. This behaviour is expected since the LGAD signal shape exhibits a larger derivative at the beginning of the pulse \cite{LGAD}. The deterioration of the performance with the threshold is reduced thanks to the time walk correction.

Finally, the time resolution after time walk correction
was extracted as a function of the position in the pad, as shown in Figure \ref{fig:resoMapB24ch3}. The bin size was chosen to ensure sufficient statistics for the computation of the time resolution. Within the statistical error, the time resolution is quite uniform.
\begin{figure}[H]
\centering

\includegraphics[width=0.5\textwidth]{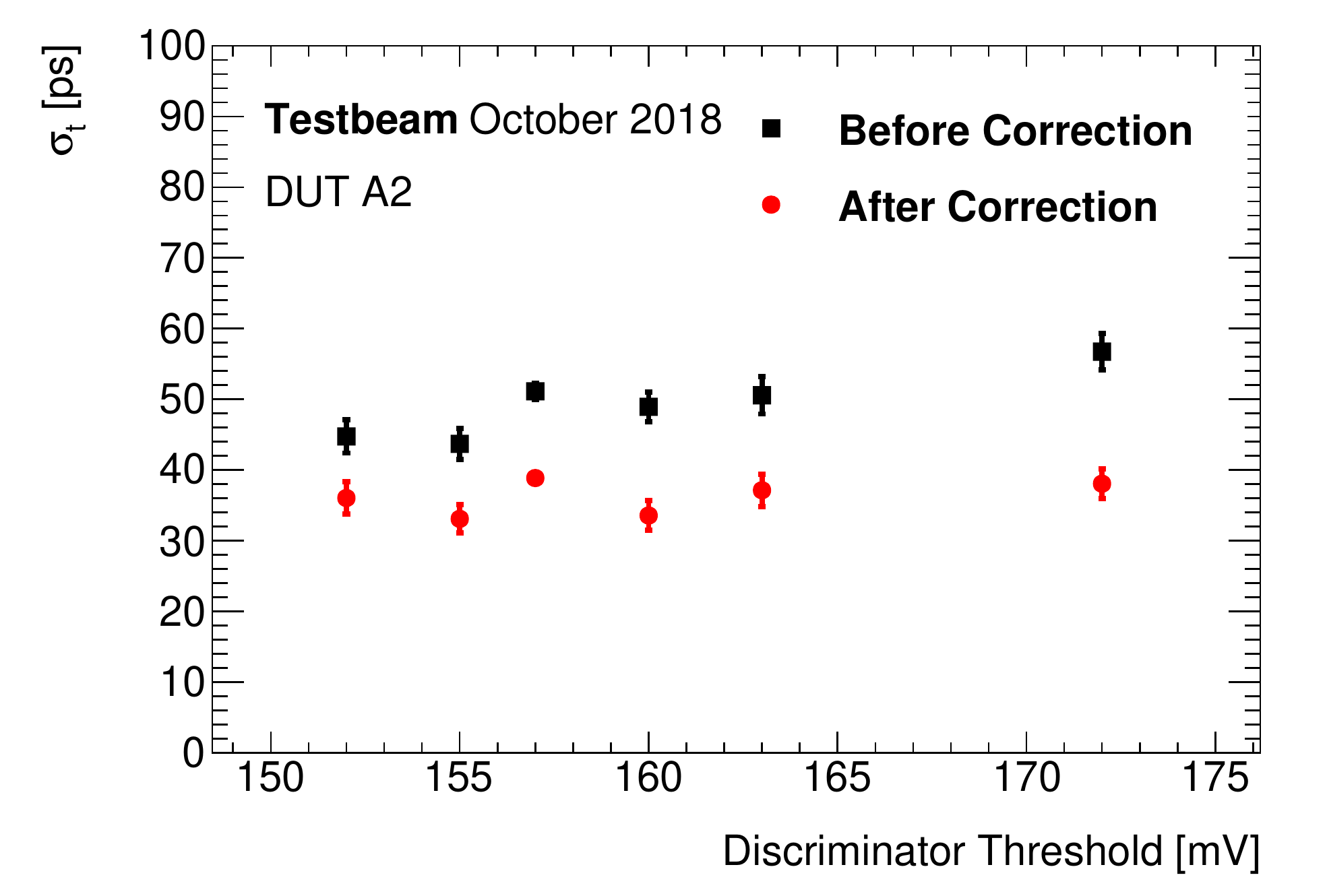}

\caption{Time resolution before and after time walk correction for a channel of ALTIROC0-LGAD bare module as a function of the discriminator threshold. A SiPM with a resolution of 40 ps is used as a time reference - it's contribution has been subtracted quadratically. The amplitude of the preamplifier probe is used to correct for the time walk. }
\label{fig:reso_DAC_B24}
\end{figure}

\begin{center}
\begin{table}[htbp]
\caption{Time resolution and the statistical error (in ps) for the 4 channels of A1 and A2.}
\centering
\begin{tabular}{|c|c|c|c|c|}
\hline

 & Ch0 & Ch1 & Ch2 & Ch3 \\ \hline
A1 & 37.9 $\pm$ 1.1 & 40.6 $\pm$ 0.9 & 43.6 $\pm$ 1.1 & 45.6 $\pm$ 1.1 \\ \hline
A2 & 36.6 $\pm$  1.1 & - & 34.7 $\pm$ 1.0 & 38.0 $\pm$ 0.9 \\

\hline

\end{tabular}

\label{tab:timeReso}
\end{table}

\end{center}

\begin{figure}[htbp]
\centering

\includegraphics[width=0.7\textwidth]{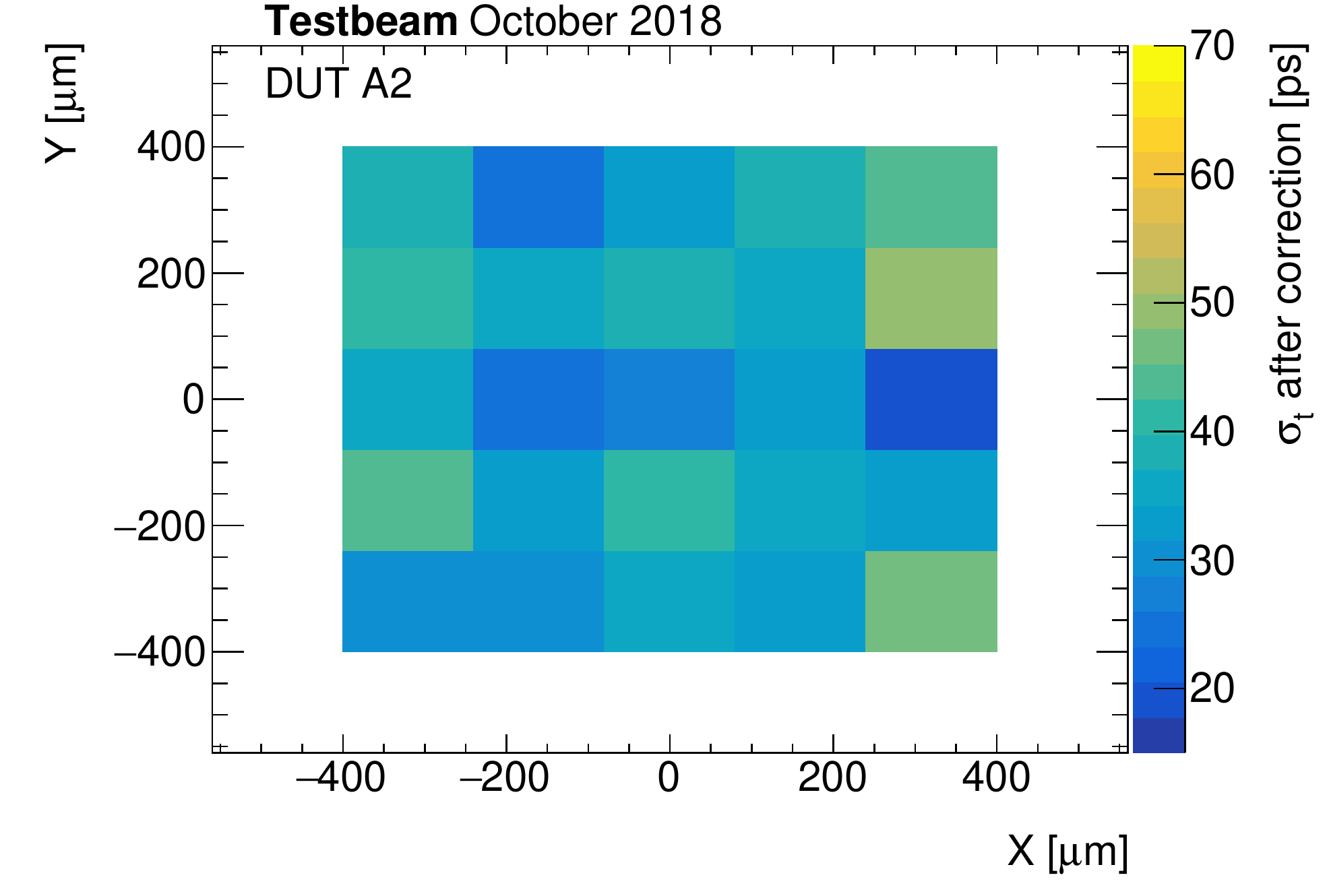}

\caption{Time resolution for a channel of an ALTIROC0-LGAD bare module as a function of the position in the pad. The time resolution has been corrected for the time walk effect using the amplitude of the preamplifier probe and the resolution of the SiPM has been subtracted. There is a minimum of 200 events in each bin (of the size of \SI{160}{\micro\metre}) so that the statistical error is about 4-5 ps.}
\label{fig:resoMapB24ch3}
\end{figure}

\end{subsubsection}

\begin{subsubsection}{Efficiency}

The efficiency map of the bare module has also been measured. The efficiency is defined as the fraction of tracks that produce a discriminator response (above a given threshold) over the total number of tracks crossing the DUT at the same position. The track is required to have a signal in the SiPM to ensure synchronicity of the telescope and waveform data. The 2D distributions of the efficiency for the 4 channels of the A1 DUT are shown in Figure \ref{fig:effMap_B23}. The discriminator threshold applied for this measurement ranges between 1.5 and 3.2 fC for the different channels. Table \ref{tab:efftable} lists the average efficiency and its statistical error for the 4 channels of A1 and A2. For the computation of the average efficiency, only the central 0.7 $\times$ 0.7 mm$^2$ bulk of the pad has been used. The bayesian approach with a beta function as a prior has been used for the calculation of the statistical error. All channels have an efficiency larger than 95\%, quite similar to the performance of the testbeam measurements of LGAD sensors mounted to simpler readout boards ~\cite{TestBeamPaper}. Within a given channel, the efficiency is constant within 1\% when varying the threshold from 1 to 9 fC.


\begin{figure}[H]
\centering
\includegraphics[width=0.7\textwidth]{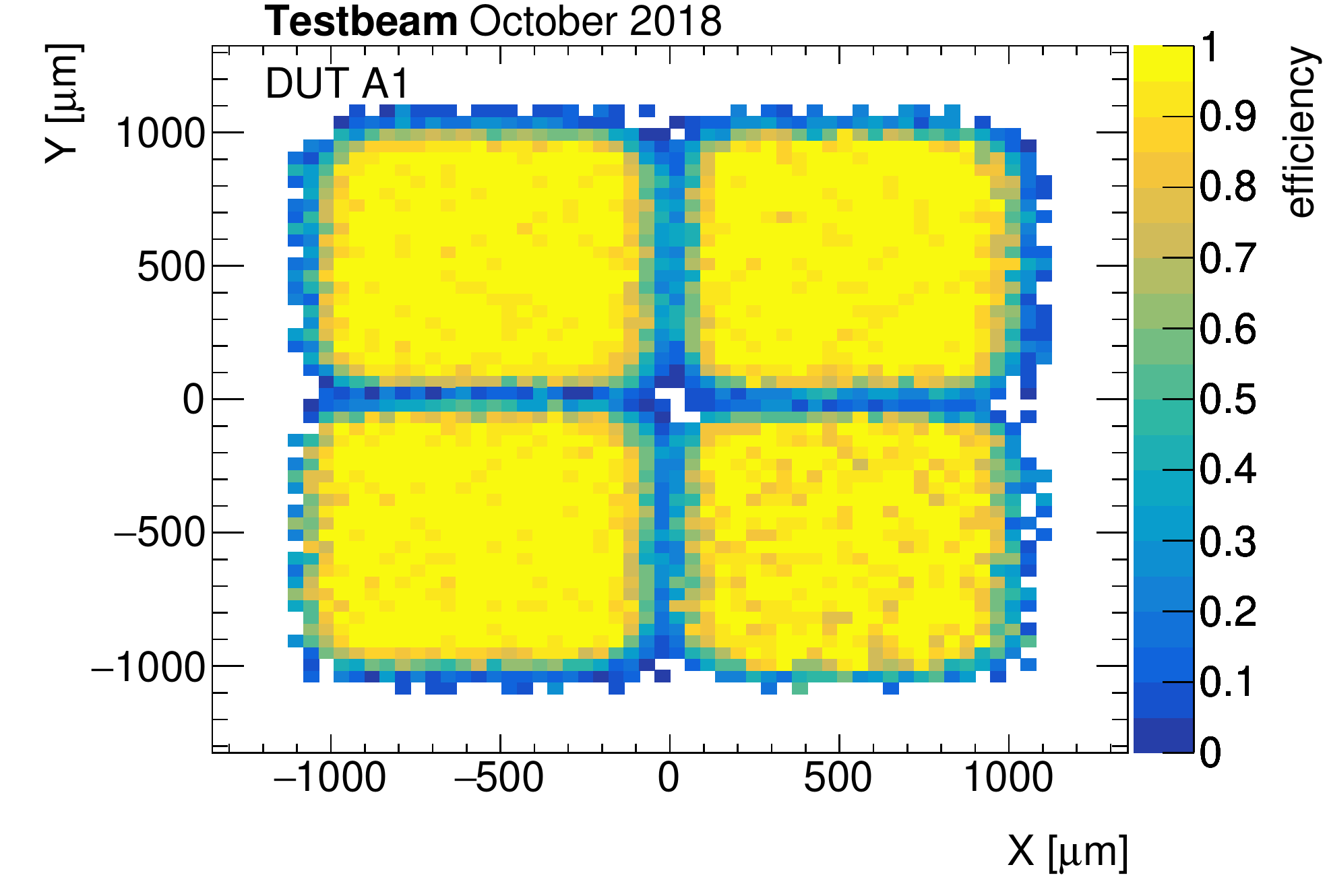}
\caption{distribution of the efficiency for the four channels of A1. 
}
\label{fig:effMap_B23}

\end{figure}

\begin{center}
\begin{table}[h]
\caption{Average efficiency (in \%) and the statistical error in the bulk  of the pad for the 4 channels of A1 and A2.}
\centering
\begin{tabular}{|c|c|c|c|c|}
\hline

 & Ch0 & Ch1 & Ch2 & Ch3 \\ \hline
A1 & 97.7 $\pm$ 0.2  & 95.2  $\pm$  0.4 & 97.6 $\pm$  0.2 & 97.4 $\pm$  0.2 \\ \hline
A2 & 97.8 $\pm$ 0.2 & - & 97.7  $\pm$  0.2 & 97.3 $\pm$  0.2 \\

\hline

\end{tabular}

\label{tab:efftable}
\end{table}

\end{center}

\end{subsubsection}

\end{subsection}

\FloatBarrier

\section{Conclusion}
\label{sec:conclusion}
An analog front-end electronics prototype for picosecond precision time measurements with LGAD sensors, named ALTIROC0, has been designed and tested with calibration signals and beam test particles. 

In calibration measurements, the various contributions to the time resolution, as well as the behaviour of the ASIC under different conditions, were studied. The jitter contribution to the time resolution, either with just the ASIC or with a module consisting of the ASIC and an LGAD sensor,  was found to be better than 20 ps for a charge larger than 5 fC. The time walk effect was corrected up to 10 ps. A 6\% improvement of the ASIC jitter for Q$_{inj}$ = 10fC was achieved during measurements at T= $-30 ^o$ C, which will be the default operating temperature for the HGTD. 

Testbeam measurements with a pion beam at CERN  were also undertaken to evaluate the performance of the module with LGAD pulses. The tested modules were operated at a bias voltage of -120 V, resulting in a most probable charge of $\sim$ 20 fC and a leakage current of O(10$^{-2}$) \SI{}{\micro\ampere}. A time resolution better than 40 ps was obtained for all channels after time walk correction, while the best-achieved performance was  34.7 $\pm$ 1 ps. This value was found to be compatible with the quadratic sum of the estimated jitter, residual of the time-walk correction and sensor contributions to the time resolution.
The time resolution was distributed uniformly in the bulk of the sensor pads and the efficiency was found to be above 95\% for all tested channels.

The resulting performance of ALTIROC0 fulfils the challenging requirements for the front-end read-out of the HGTD at the HL-LHC. The next iteration of the ASIC, ALTIROC1, will introduce the digital part of the front-end readout. It will integrate 25 channels, including in each two Time-to-Digital converters followed by an SRAM. Along with the characterisation of the digital part of the readout chain, the new iteration will be evaluated under various irradiation conditions and at the limits of its dynamic range.


\acknowledgments
We acknowledge CERN for the very successful operation of the SPS and thank the North Area test beam support team. We also thank the HGTD community for their valuable inputs and discussions.


\begin{thebibliography}{99}
\bibitem{ATLAS} ATLAS collaboration, \emph{The ATLAS Experiment at the CERN Large Hadron Collider}, \href{http://dx.doi.org/10.1088/1748-0221/3/08/S08003}{JINST \textbf{3}
(2008) S08003} 

\bibitem{ITk} ATLAS collaboration, \emph{Technical Design Report for the ATLAS Inner Tracker Pixel Detector}, \href{https://cds.cern.ch/record/2285585}{CERN-LHCC-2017-021}     

\bibitem{LGAD} H. F. W. Sadrozinski et al., \emph{Ultra-fast silicon detectors (UFSD)}, \href{https://www.sciencedirect.com/science/article/pii/S0168900216301279}{Nucl. Instrum. Meth. \textbf{A831} (2016) 18}

\bibitem{HGTD-TP} ATLAS collaboration, \emph{Technical Proposal: A High-Granularity Timing Detector for the ATLAS Phase-II upgrade}, \href{http://cds.cern.ch/record/2623663}{CERN-LHCC-2018-023}


\bibitem{ALT0} 
C. De La Taille et al., \emph{ALTIROC0, a 20 pico-second time resolution ASIC for the ATLAS High Granularity Timing Detector (HGTD)} in \emph{Proceedings of the Topical Workshop on Electronics for Particle Physics (TWEPP-17)}, Santa Cruz, California, U.S.A, September 2017,
\href{https://pos.sissa.it/313/006}{PoS TWEPP-17 (2018) \textbf{006}}



\bibitem{nicolo} Cartiglia et al, \emph{Beam test results of 16 ps timing system based on ultra-fast silicon detectors}, \href{https://www.sciencedirect.com/science/article/pii/S0168900217300219?via\%3Dihub}{Nucl. Instrum. Meth \textbf{A850} (2017)}  \href{https://arxiv.org/abs/1608.08681}{[1608.08681]}

\bibitem{TestBeamPaper} L. Masetti et al, \emph{Beam test measurements of Low Gain Avalanche Detector single pads and arrays for the ATLAS High Granularity Timing Detector}, \href{https://iopscience.iop.org/article/10.1088/1748-0221/13/06/P06017}{JINST \textbf{13} (2018)} \href{https://arxiv.org/abs/1804.00622}{[1804.00622]}

\bibitem{telescope}
H. Jansen et al., \emph{Performance of the EUDET-type beam telescopes}, \href{https://epjtechniquesandinstrumentation.springeropen.com/articles/10.1140/epjti/s40485-016-0033-2}{EPJ Techniques and Instrumentation \textbf{3} (2016) 7} \href{https://arxiv.org/abs/1603.09669}{[1603.09669]}

\bibitem{FEI4} J. Albert et al., \emph{Prototype ATLAS IBL Modules using the FE-I4A Front-End Readout Chip}, \href{https://iopscience.iop.org/article/10.1088/1748-0221/7/11/P11010}{JINST \textbf{7} (2012) P11010} \href{https://arxiv.org/abs/1209.1906}{[1209.1906]}

%
%
%
%
%
%
%
\end{thebibliography}
\end{document}